\documentclass[twocolumn]{aastex62}

\usepackage[T1]{fontenc}
\usepackage{natbib}
\usepackage{amsmath}
\usepackage{xcolor}
\usepackage[normalem]{ulem}  
\usepackage{comment}

\hypersetup{linkcolor=blue,citecolor=blue,filecolor=blue,urlcolor=blue}

\newcommand{\BEL}{\textsc{bel}}
\newcommand{\BLR}{\textsc{blr}}
\newcommand{\LIL}{\textsc{lil}}
\newcommand{\HIL}{\textsc{hil}}
\newcommand{\AGN}{\textsc{agn}}
\newcommand{\EW}{\textsc{ew}}
\newcommand{\SED}{\textsc{sed}}
\newcommand{\UV}{\textsc{uv}}
\newcommand{\EUV}{\textsc{euv}}
\newcommand{\xray}{\textsc{x}ray}
\newcommand{\LINER}{\textsc{liner}}
\newcommand{\WLQ}{\textsc{wlq}}
\newcommand{\TR}{\textsc{tr}}
\newcommand{\HeII}{He\textsc{ii}}

\newcommand{\CIV}{C\textsc{iv}}
\newcommand{\MgII}{Mg\textsc{ii}}
\newcommand{\Hbeta}{H{\footnotesize $\beta$}}

\graphicspath{{./}{figures/}}

\shorttitle{Broad-line phenomenology unified through radiative filtering}
\shortauthors{Naddaf et al.}

\begin{document}

\title{Radiative filtering unifies broad-line phenomenology in active galactic nuclei}

\correspondingauthor{Mohammad Hassan Naddaf}
\email{mh.naddaf@uliege.be}
   
\author[0000-0002-7604-9594]{Mohammad Hassan Naddaf}
\affil{Institut d'Astrophysique et de Géophysique, Université de Liège, Allée du six août 19c, B-4000 Liège (Sart-Tilman), Belgium}

\begin{abstract}
Broad emission lines (\BEL{}s) are a defining feature of active galactic nuclei (\AGN{}s), yet they weaken or disappear in both very low- and very high-accretion systems. These regimes are typically treated separately, and a unified physical explanation has remained elusive. Here we show that this behavior arises if line formation is governed not by the intrinsic luminosity of the central engine, but by the ionizing radiation field that survives filtering before reaching the broad-line region (\BLR{}). In this picture, line production depends on the product of intrinsic ionizing capability and an effective transmission. Because the former increases from low accretion rates while the latter declines at high accretion rates, the effective ionizing field naturally develops a finite and non-universal window for \BEL{} formation. This framework unifies the absence or extreme faintness of \BEL{}s in low-luminosity \AGN{}s, \LINER{}s, and weak-line quasars (\WLQ{}s), and accounts for the Baldwin effect and the $R_{\rm Fe}$ trend. It also necessarily implies the breakdown of standard \BLR{}-based scaling relations in extreme accretion regimes. We show that a minimal quantitative realization reproduces this behavior across black-hole mass, accretion rate, and radiative efficiency. These results suggest that \AGN{} emission-line phenomenology is governed by global regulation of the ionizing radiation field rather than by mere presence or condition of local gas.
\end{abstract}


\section{Introduction} \label{sec:intro}

\BEL{}s are among the defining observational signatures of \AGN{}s, yet they are not universally present. \BEL{}s arising from the virialized \BLR{} are observed to weaken or disappear in two apparently opposite regimes: at very low accretion rates, such as in low-luminosity \AGN{}s and \LINER{}s \citep{elitzur2009, cao2010}, and at very high accretion rates, as seen in \WLQ{}s \citep{shemmer2010, luo2013,luo2015}. These two regimes are usually attributed to distinct physical mechanisms, despite both marking a departure from the usual broad-line behavior.

At the same time, both theory and observations indicate that the radiation field incident on the \BLR{} need not coincide with the intrinsic emission of the central engine. Accretion flows approaching the Eddington limit are expected to develop strong outflows and geometrically thick inner structures that can redirect or attenuate ionizing radiation before it reaches larger radii \citep{abramowicz1988, murray1995, proga2000, wang_shielding2014}. Observationally, high-accretion systems frequently show signatures of dense outflows and \xray{} weakness, requiring a reduced ionizing flux reaching the line-emitting gas \citep{luo2013, luo2015}. More generally, global \BLR{} models such as the Failed Radiatively Accelerated Dusty Outflow (FRADO) scenario indicate that line-emitting structure and illumination are regulated by global parameters including accretion rate and black-hole mass \citep{czerny2011,naddaf2021,naddaf2025L}. This is also reflected in the quasar main sequence, where \BEL{} properties correlate primarily with Eddington ratio and secondarily with black-hole mass \citep{naddaf2025L, zamfir2010}. Moreover, the systematic weakening of emission-line equivalent widths (\EW{}s) with increasing luminosity, known as Baldwin effect, further indicates that the effective ionizing field governing line formation does not scale trivially with the observed continuum \citep{baldwin1977}.

These considerations require a shift in perspective. The relevant question is not whether gas is present, but whether the ionizing radiation reaching it exceeds the threshold for line production. We show that \BEL{} appearance is governed by a transmission-regulated (\TR{}) condition where the effective ionizing field is the product of intrinsic photon production and a filtering-induced transmission. Since photon production rises away from low accretion states, while transmission declines once filtering becomes significant, the effective ionizing field naturally develops a bounded maximum. A finite line-production window therefore follows generically. This accords with earlier results showing that the \BLR{} may see a modified or filtered ionizing continuum rather than the same spectral energy distribution (\SED{}) seen by the observer \citep[e.g.,][]{korista1997, ferland2020}.

This principle does not depend on the filtering microphysics, nor on a unique mathematical form for the transmission function. It is a generic consequence of combining a rising ionizing capability at low accretion rates with significant suppression at high accretion rates.

The paper is organized as follows. Section~\ref{sec:structure} introduces the conceptual framework. Section~\ref{sec:method} presents a minimal method of quantitative realization. Section~\ref{sec:results} shows the resulting finite and non-universal \Hbeta{} window. Section~\ref{sec:discussion} discusses the implications for broad-line phenomenology. Section~\ref{sec:conclusion} summarizes the conclusions.

\section{Conceptual framework}\label{sec:structure}

\BEL{} visibility is governed by the ionizing radiation reaching the \BLR, rather than by the intrinsic luminosity of the central engine alone \citep{korista1997, ferland2020}. We formulate this as a \TR{} framework.

\subsection{Transmission-regulated line-production window}

The effect of filtering can be represented through an effective transmission factor $T_{\rm net}$, defined as the fraction of ionizing radiation that reaches the line-emitting gas. This quantity is intended to capture the net effect of all processes that reduce the incident ionizing field, without explicitly modeling the detailed radiative-transfer or microphysics of the intervening medium.

The appearance of \BEL{}s is governed by the condition
\begin{equation}
\Phi_{\rm eff} = \Phi_{\rm int} \, T_{\rm net}~\geq~ \Phi_{\mathrm{\BEL{}, min}},
\end{equation}
or equivalently, $\varphi_{\rm \BEL{}} \, T_{\rm net} \geq 1$ where $\varphi_{\rm \BEL{}}$ is the dimensionless ratio of ionizing photon flux to the minimum flux required for \BEL{} production. This relation defines the central operative condition of the framework.

Since the ionization parameter is defined as
\begin{equation}
U = {\Phi_{\rm int}}/({n_{\rm H} \,c}),
\end{equation}
it directly implies the reduction in ionization level as
\begin{equation}
U_{\rm eff} = U_{\mathrm{int}}\, T_{\rm net},
\end{equation}
which holds if one compares two states of the same $n_{\rm H}$ with and without assumption of radiative filtering.

The key point is qualitative and general. At low accretion rates, the intrinsic ionizing flux is insufficient; $\varphi_{\rm \BEL{}}$ remains below the threshold. As the accretion rate increases, the ionizing capability rises, while filtering increasingly counteracts its transmission; once filtering dominates, the transmitted ionizing field decreases even if intrinsic photon production continues to grow. The product $\varphi_{\rm \BEL{}}T_{\rm net}$ thus develops a bounded maximum, so line production is confined to a finite region of parameter space rather than increasing monotonically with accretion rate. This behavior is generic provided that filtering suppresses the transmitted ionizing field strongly enough that $\varphi_{\rm \BEL{}}T_{\rm net}$ declines at high accretion.

\subsection{Effective ionization across global parameter space}

The transmission as a general bounded function is
\begin{equation}
\label{eq:transmission}
T_{\rm net}(\lambda, \,t) \equiv
f\left(\dot{m}, M_\bullet, \eta\,;~ \mathcal{V}_{t}\right),
\end{equation}
where $\dot{m}$ is the dimensionless accretion rate characterizing the global or time-averaged accretion state, $M_\bullet$ is the black-hole mass, $\eta$ is the radiative efficiency, and $\mathcal{V}_{t}$ denotes short-timescale variability in the continuum and/or filtering structure. The function $f$ encodes geometrical and microphysical changes in the filtering structure and determines the wavelength-dependent ionizing continuum transmitted to the line-emitting gas.

Observed \UV{}/optical luminosities can vary on timescales much shorter than those associated with secular changes in the mass supply through the disk. Such variability may arise from thermal fluctuations, disk-corona coupling, \textsc{mhd}  turbulence, or changes in the filtering structure. Therefore, $\dot{m}$ should not be identified with the instantaneous luminosity inferred from a single-epoch spectrum. In this Letter, we focus on the population-level and time-averaged consequences of the framework. Equivalently, the calculations below correspond to fixed or averaged $\mathcal{V}$, so that the illustrative transmission is treated as a function of the global parameters.

\begin{figure*}[t]
\centering
\includegraphics[width=1.9\columnwidth]{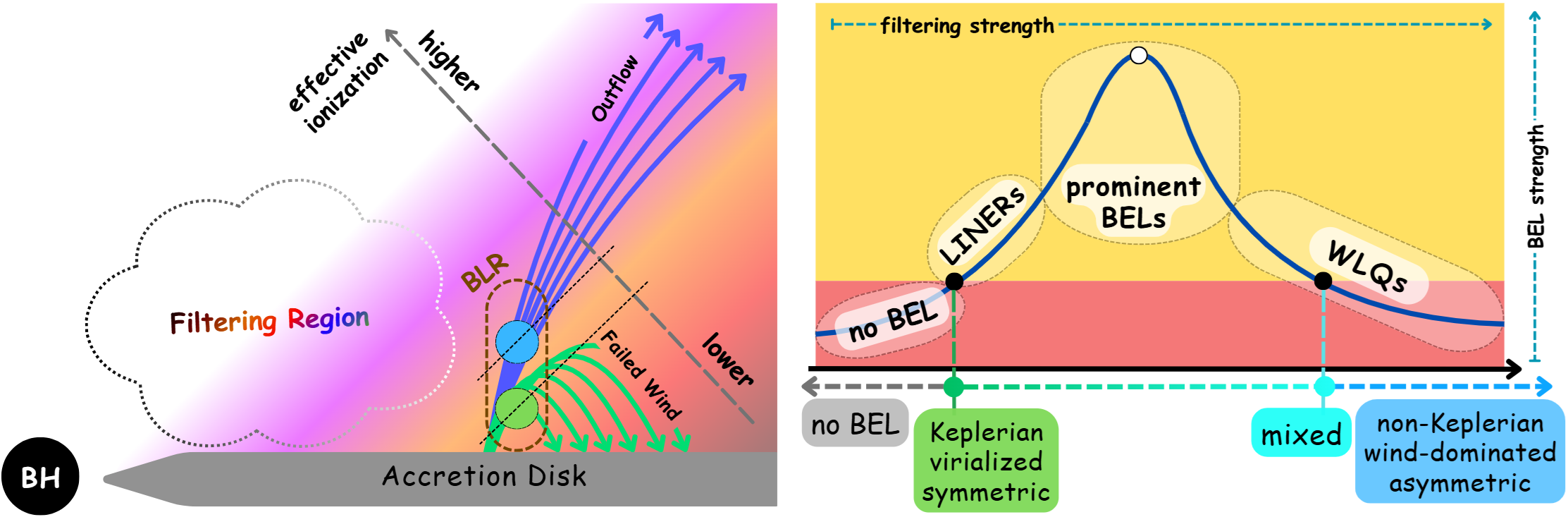}
\caption{Conceptual picture of \TR{} \BEL{} formation.
Left: ionizing radiation from the central engine is filtered before reaching the \BLR{}, so the line-emitting gas responds to the transmitted rather than the intrinsic ionizing field. In this schematic, different regions within the \BLR{} may be characterized by distinct dynamical regimes. Regions closer to the equatorial plane are expected to be more virialized (green circle), while higher-altitude or outflowing regions may be dominated by non-virial motions (blue circle). Black dotted lines indicate different effective ionization layers, reflecting anisotropic illumination. As the effective ionizing field is reduced by filtering, the subset of gas that satisfies the line-production condition may shift between these regions. This provides a natural interpretation for changes in line profile morphology, including increased blueshifts and wind signatures. It may also lead to shorter effective reverberation lags if the dominant emitting region shifts to outflowing gas (blue circle) which is geometrically favored for a shorter light-travel path to the observer, even if located at larger physical radii.
Right: Schematic placement of different \AGN{} populations along this sequence, including systems with no detectable \BEL{}s, \LINER{}s, typical broad-line \AGN{}s, and \WLQ{}s. In this framework, the appearance, shape, and strength of \BEL{}s are governed by the degree to which the effective ionizing flux incident on the gas exceeds the threshold required for line production.}
\label{fig:location}
\end{figure*}

Combining this transmission with the line-formation condition yields a finite window in $(\dot{m}, M_\bullet, \eta)$ space within which \BEL{}s can be produced. This window constrains the conditions for observable line formation, not the existence of \BLR{} gas itself. This qualitative, generic behavior is illustrated schematically in Fig.~\ref{fig:location}, where the effective ionizing field is shown to rise from low accretion rates and then decline once filtering becomes dominant. An illustrative realization is presented in Methodology.

\subsection{\BEL{}: illumination not mere gas presence}

The \TR{} picture thus constrains the conditions for producing observable \BEL{}s, rather than the existence of \BLR{} gas; it may be present outside the line-production window, but if the transmitted ionizing flux falls below the threshold required for efficient excitation, no detectable \BEL{} will be produced.
This distinction is schematically shown in Fig.~\ref{fig:location}. Filtering reduces the effective ionizing illumination of \BLR{} thereby regulates the ionization state and the resulting \BEL{} properties. The appearance of \BEL{}s is thus controlled by the ionizing radiation field reaching the gas, rather than by mere presence of gas.

The framework therefore separates two questions that are often conflated: whether line-emitting material exists, and whether it is sufficiently illuminated to produce an observable \BEL{}. This distinction becomes essential in extreme accretion regimes, where \BEL{} weakness need not imply the disappearance of the \BLR{} itself.

This model is orthogonal to orientation-based unification. Geometric orientation modulates the observed continuum and \BEL{} profiles \citep{antonucci1993}, but the \TR{} condition governs whether the ionizing field reaching the \BLR{} is sufficient for \BEL{} production.

\section{Quantitative realization}\label{sec:method}
We now present a minimal method for quantitative realization of the \TR{} framework.

\subsection{Transmission prescription}
We describe the effect of filtering of the ionizing radiation field through an effective transmission factor $T_{\rm net}$, defined as the fraction of ionizing photons that reach the line-emitting gas. This quantity encapsulates the net impact of all processes that reduce the ionizing field incident on the \BLR{}, without explicit treatment of the detailed microphysics.

The transmission factor is modeled as a bounded, monotonic function of global \AGN{} parameters.
For a generic variable $X$, we adopt the form
\begin{equation}
T(X)=\frac{1}{\exp\left[k\left(\ln X-\ln X_0\right)\right]+1},
\end{equation}
where $X_0$ defines the characteristic transition scale and $k$ sets the slope.

This choice is best regarded as a minimal illustrative realization rather than a special, unique, or exclusive model. The existence of a finite line-production window does not depend on the specific mathematical form adopted for the function $T_{\rm net}$. It requires only that the transmitted ionizing flux remain unchanged at low filtering and eventually decreases sufficiently at high filtering.

\begin{figure*}[t]
\centering
\includegraphics[width=0.85\textwidth]{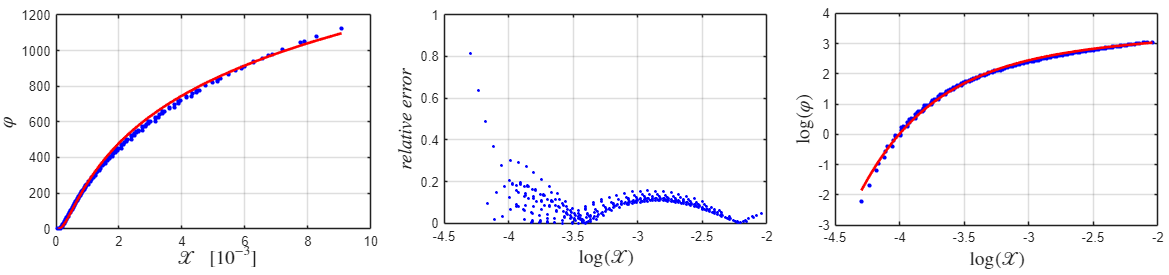}
\caption{Accuracy of the analytic approximation for the ionizing photon-flux ratio. Comparison between the numerically evaluated $\varphi_{\mathrm{H}\beta}$ and the analytic approximation as a function of $\mathcal{X}$. Left: $\varphi_{\mathrm{H}\beta}(\mathcal{X})$ on linear axes. Middle: relative deviation between numerical and analytic values. Right: comparison in log space, showing agreement over several decades in $\mathcal{X}$. The analytic expression reproduces the numerical result across the explored parameter space, with typical deviations below $\sim 20\%$.}\label{fig:approximation}
\end{figure*}

\subsection{Ionizing photon flux and dimensionless ratio}

We now specialize to \Hbeta{} as a representative \BEL{}. 
The intrinsic ionizing capability of the source is quantified by the dimensionless photon-flux ratio
\begin{equation}
\varphi_{\mathrm{H}\beta}=\frac{\Phi_{\rm H}}{\Phi_{\rm H\beta,\min}},
\end{equation}
where $\Phi_{\rm H}$ is the ionizing photon flux at the characteristic \BLR{} radius, $\Phi_{\rm H\beta,\min}\simeq3\times10^{17}\,{\rm s^{-1}\,cm^{-2}}$ is the minimum flux required for H{\footnotesize $\beta$} production.

Photoionization calculations show that broad \Hbeta{} emission occurs only over a restricted region of the density--ionizing-flux plane \citep{korista1997a, korista2004}. 
It is shown that gas exposed to very low ionizing photon fluxes,
$\log \Phi_{\rm H} \lesssim 17.5$
can still emit optical recombination lines.
This is also consistent with a simple estimate from the
definition of the ionization parameter.
For characteristic low-ionization lines (\LIL{s}) \BLR{} conditions,
\begin{equation}
\Phi_{\rm H}
\simeq 3.0\times10^{17}
\left(\frac{U}{10^{-2}}\right)
\left(\frac{n_{\rm H}}{10^{9}\ {\rm cm^{-3}}}\right)
{\rm s^{-1}\ cm^{-2}}.
\end{equation}

It should be however note that the quantity $\Phi_{{\rm H}\beta,\min}$ is not intended as a sharp physical threshold for hydrogen recombination. It is used here as a lower practical boundary for responsive broad \Hbeta{} emission which actually serves as a normalization for $\varphi_{\rm H\beta}$. Changing its value would shift the numerical threshold crossings in $\dot m$, but would not change the existence of a finite line-production window. Thus the adopted value should be regarded as an illustrative fiducial scale rather than a universal hard boundary.

The ionizing flux $\Phi_{\rm H}$ implicitly accounts for anisotropy and geometric dilution of the radiation field, defined as
\begin{equation}
\Phi_{\rm H}=f_{\rm ans} \frac{Q_{\rm H}}{4\pi R_{\rm \BLR{}}^2},
\end{equation}
where $Q_{\rm H}$ is the photon production rate above the Lyman limit.
The continuum is modeled using a multicolor blackbody corresponding to a standard thin accretion disk, which provides an approximation for the scaling of ionizing photon production. The factor $f_{\rm ans}\simeq0.1$ encapsulates deviations from isotropic illumination. In the illustrative implementation presented here, we adopt $f_{\rm ans}\sim0.1$, consistent with a modestly anisotropic radiation field \citep{naddaf2025}. The factor $f_{\rm ans}$ mainly sets the normalization: it does not alter the existence of a bounded maximum in the illustrative realization, but it does shift the threshold crossings and hence the boundaries of the line-production window.

The characteristic \BLR{} radius can be estimated as
\begin{equation}
R^3_{\rm \BLR{}}\propto \dot{m}\, M_\bullet^2\,\eta^{-1},
\end{equation}
from the radial profile of disk effective temperature.

The dependence of $\varphi_{\mathrm{H}\beta}$ on global parameters is computed numerically and approximated then to
\begin{equation}\label{eq:phi_fit}
\varphi_{\mathrm{H}\beta}\simeq3.4\times10^3\,\mathcal{X}^{\alpha}~
\exp[-1/\left(160\,{\mathcal{X}^{\beta}}\right)],
\end{equation}
where $\mathcal{X}=\eta^{0.5}\dot{m}^{0.3}M_\bullet^{-0.3}$, $\alpha=1/5$, and $\beta=3/4$. The analytic approximation reproduces the numerical evaluation across the explored parameter space, with typical deviations below $\sim 20\%$ (Fig.~\ref{fig:approximation}).

Equation~(\ref{eq:phi_fit}) is an internal fit to the adopted thin-disk calculation and should be interpreted only as an illustrative trend when the Eddington ratio is large enough for slim-disk effects to matter.

\subsection{Coupling condition for line formation}
Broad H{\footnotesize $\beta$} emission occurs when the transmitted ionizing photon flux exceeds the minimum required for efficient line production. This defines the condition
\begin{equation}
\varphi_{\mathrm{H}\beta} \, T_{\rm net}\geq1,
\end{equation}
which couples the intrinsic ionizing capability of the source to the effective transmission of the radiation field.

Filtering is treated as a reduction in the effective ionizing flux incident on the gas, not geometric obscuration.

\section{Results}\label{sec:results}
For the purpose of illustration, we evaluate the \Hbeta{} production window over representative ranges of global parameters motivated by observed \AGN{} populations:
\begin{equation}
\nonumber
\log\dot{m}\in[-4,2],~~~
\log M_\bullet\in[6,11],~~~
\eta\in[0.038,0.32].
\end{equation}

These ranges are not intrinsic to the formulation. The variables themselves are not restricted to finite intervals, and the qualitative behavior of the model does not depend on the specific domain over which it is evaluated. The window is a structural consequence of the formulation and is therefore independent of the specific parameter ranges adopted for illustration.

\begin{figure*}[t]
\centering
\includegraphics[width=0.95\textwidth]{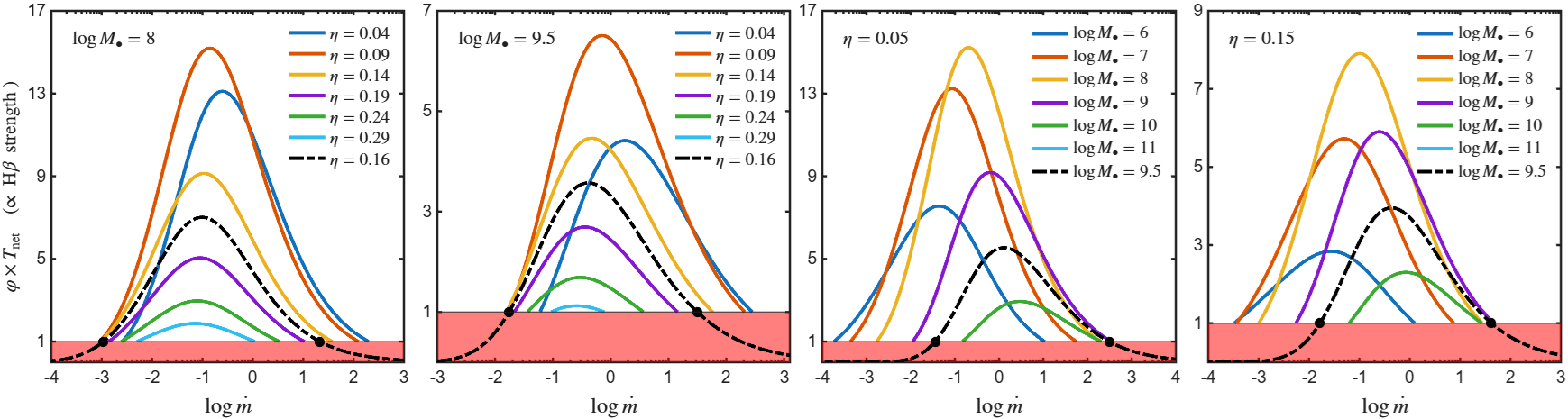}
\caption{Illustrative emergence of a finite H{\footnotesize $\beta$} production window. The effective line-forming quantity $\varphi_{\mathrm{H}\beta}\,T_{\rm net}$ (proportional to H{\footnotesize $\beta$} strength) is shown as a function of $\log\dot{m}$ for representative values of black-hole mass $M_\bullet$ and radiative efficiency $\eta$. In each panel, one parameter is held fixed while the other is varied, as indicated. The intrinsic ionizing capability increases from low accretion rates, whereas transmission decreases once filtering becomes significant, resulting in a bounded maximum. The two intersections with the threshold condition $\varphi_{\mathrm{H}\beta}\,T_{\rm net}=1$ (horizontal line) define the lower and upper boundaries of the allowed broad H{\footnotesize $\beta$} window. Black points mark these intersections for a representative example (dashed curve), illustrating the minimum and maximum $\log\dot{m}$ that satisfy the line-production condition. This behavior is generic to filtering formalism and is illustrated here with the sigmoid realization adopted in this work.}
\label{fig:hbeta_window}
\end{figure*}

The net transmission is taken as first-order separable,
\begin{equation}
T_{\rm net}=T(\dot{m}) \, T(M_\bullet) \, T(\eta),
\end{equation}
approximating the independent influence of each parameter on the effective ionizing field (higher-order couplings are not included).
The adopted functional forms are
\begin{align}
&T(\eta)&&=1/\left(\exp\left[3.1\left(\log\eta+1\right)\right]+1\right),\\
&T(\dot{m})&&=1/\left(\exp\left[0.72\left(\log\dot{m}+1\right)\right]+1\right),\\
&T(M_\bullet)&&=1/\left(\exp\left[0.58\left(8.5-\log M_\bullet \right)\right]+1\right).
\end{align}

The coefficients are fixed by mapping the observed dynamic ranges of the parameters into comparable transition intervals and are not tuned to reproduce emission-line data.
The adopted sigmoid forms span the full domain of the variables, and the choice of slopes and transition scales is made only to place the transitions within observationally relevant regimes, without affecting the qualitative behavior. Within the adopted illustrative realization, the window is structural and not an artifact of the plotted parameter bounds.

\subsection{Realization of line-production window}
Figure~\ref{fig:hbeta_window} shows the resulting behavior for representative values of $M_\bullet$ and $\eta$. The effective ionizing field exhibits a well-defined peak at intermediate accretion rates. The two crossings of the condition $\varphi_{\mathrm{H}\beta}\, T_{\rm net}=1$ on either side of this peak define the lower and upper boundaries of the H{\footnotesize $\beta$} production window.

For fixed $M_\bullet$ and $\eta$, \BEL{} first appears above a minimum accretion rate, reaches a maximum at intermediate values, and then weakens again at higher accretion rates owing to increased filtering. The bounded maximum arises because increases in intrinsic ionizing capability are eventually offset by enhanced filtering, preventing a monotonic growth of the effective ionizing field.

H{\footnotesize $\beta$} \BEL{} therefore occupies only a bounded range of accretion rate, bounded at low $\dot{m}$ by insufficient ionizing photon production and at high $\dot{m}$ by reduced transmission. The two boundaries arise from the same condition, implying a single underlying regulatory mechanism.

The dependence on $M_\bullet$ is not monotonic: changing $M_\bullet$ can either increase or decrease the effective ionizing field and shift the window, depending on the location in parameter space. Variations in $\eta$ similarly modulate both the onset and the suppression of line production across parameter space. As a result, the position and extent of the allowed H{\footnotesize $\beta$} window vary jointly with $M_\bullet$ and $\eta$. Therefore, sources with comparable nominal Eddington ratios can exhibit substantially different \BEL{} properties depending on their global parameters.

\begin{figure*}[t]
\centering
\includegraphics[width=0.95\textwidth]{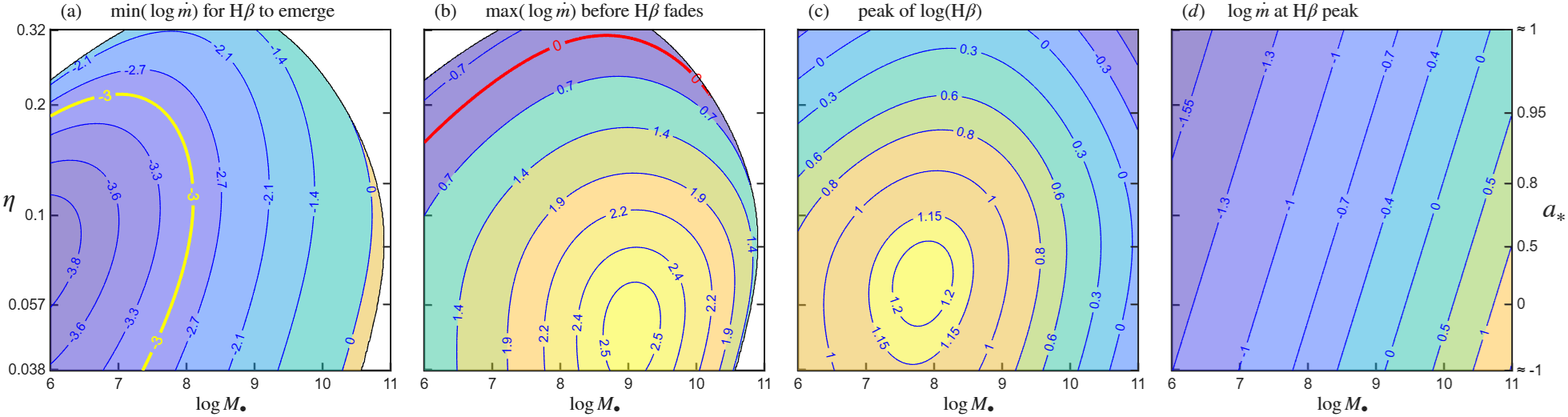}
\caption{Non-universal H{\footnotesize $\beta$} production window across global parameter space. Each panel summarizes a different aspect of the condition $\varphi_{\rm H\beta}\,T_{\rm net} \ge 1$ in the $(\log M_\bullet,\eta)$ plane, with spin $a_\ast$ shown as the corresponding proxy on the right-hand axis. (a) Minimum $\log\dot{m}$ required for H{\footnotesize $\beta$} to emerge. (b) Maximum $\log\dot{m}$ before H{\footnotesize $\beta$} fades as the transmitted ionizing field becomes insufficient. (c) Peak of $\log$($\varphi_{\mathrm{H}\beta}\, T_{\rm net}$), proportional to the peak of H{\footnotesize $\beta$} strength. (d) $\log\dot{m}$ at H{\footnotesize $\beta$} peak. The white regions mark parameter combinations for which the threshold condition is not satisfied. The figure shows explicitly that neither the onset nor the disappearance of \BEL{}s is described by a universal accretion-rate boundary.}
\label{fig:contours}
\end{figure*}

\subsection{Non-universality of \BEL{} production window}

The position of this window varies systematically across parameter space. Figure~\ref{fig:contours} maps the lower and upper limits of H{\footnotesize $\beta$} production in the $(\log M_\bullet,\eta)$ plane and shows that neither boundary is universal.

The minimum $\dot{m}$ required for H{\footnotesize $\beta$} emergence depends jointly on $M_\bullet$ and $\eta$. The upper boundary, beyond which the transmitted ionizing field again becomes insufficient for line formation, shows the same parameter dependence. There is thus no single critical accretion rate to separate sources with and without \BEL{}s.

The effective line-forming quantity $\varphi_{\mathrm{H}\beta}\, T_{\rm net}$ reaches a maximum in parameter space where photon production and filtering are optimally balanced. The global maximum (Fig.~\ref{fig:contours}c) does not follow trivially from the centers of the individual transmission functions, but instead emerges from their combined effect.

\section{Discussion}\label{sec:discussion}

The central implication of this work is that \BEL{} appearance can be understood as a threshold phenomenon in the effective ionizing radiation field reaching the \BLR{}. The key qualitative consequence is a bounded maximum in this effective field, which yields a finite line-production window. This conclusion does not depend on the detailed microphysics of the filtering medium, or on a unique transmission law.

\subsection{Possible physical realization of the filtering layer}

The filtering medium should not be interpreted as a passive sink of ionizing photons. Any reduction of the ionizing continuum incident on the \BLR{} must be accompanied by redistribution of the same energy through scattering, reprocessing, anisotropic escape, mechanical work, or other channels. Thus, $T_{\rm net}$ is not an energy-loss term, but an effective parametrization of the ionizing radiation field incident on the classical \BLR{}.

A plausible physical realization is a dense, high-column, stratified inner disk atmosphere or wind-base structure located between the compact ionizing source and at least part of the \BLR{}. In such a structure, the illuminated surface may be highly ionized, while deeper layers are denser and more optically thick. The surface layers can scatter, redirect, and partially transmit ionizing radiation, whereas photons that penetrate into deeper or less ionized layers are more efficiently absorbed and reprocessed, especially in the \EUV{} and soft-\xray{} bands. The continuum reaching the \BLR{} can thus be reduced and reshaped, rather than simply dimmed.

\begin{figure*}[t]
\centering
\includegraphics[width=0.7\textwidth]{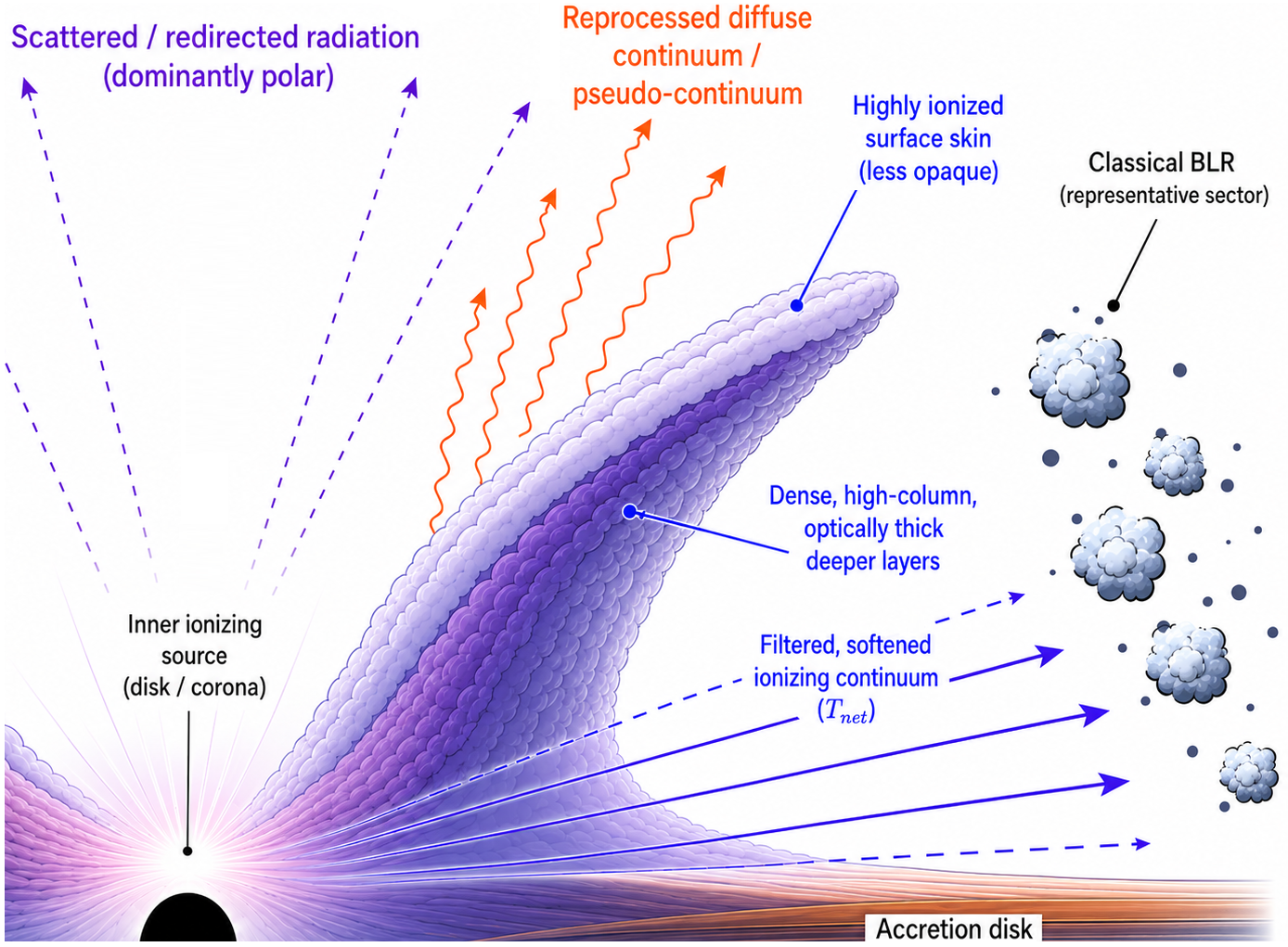}
\caption{Conceptual illustration of one possible physical realization of the filtering layer in the strong-filtering regime. The filtering medium is shown as a stratified inner disk atmosphere or wind base, physically connected to the accretion flow. Its highly ionized surface can scatter and redirect part of the ionizing radiation, while denser, high-column, optically thick deeper layers can absorb and reprocess part of the incident power into diffuse continuum or pseudo-continuum emission. The classical \BLR{} receives a reduced and reshaped ionizing continuum, represented by $T_{\rm net}$, through less opaque surface or angular channels. This schematic is not to scale, not a radiative-transfer model, and not intended to represent a unique geometry for all \AGN{}s.}
\label{fig:possible_filter}
\end{figure*}

Figure~\ref{fig:possible_filter} illustrates a possible realization of this structure in the strong-filtering regime. The broader \TR{} formalism does not require a unique geometry; it only requires that the ionizing continuum incident on the classical \BLR{} can differ from the intrinsic continuum emitted by the inner source. In the example shown, the surviving ionizing continuum reaches the line-emitting gas mainly through lower-opacity surface layers or angular channels where the effective column density is smaller. The figure is intended only as a conceptual illustration of one possible filtering geometry.

This stratified picture addresses the energy-budget constraint as well. If part of the filtering layer is only moderately ionized, it may emit bound-bound lines, free-bound recombination continua, and free-free continuum radiation. However, in the dense, high-column regime considered here, large optical depths, collisional suppression, velocity shear, and thermalization can prevent this emission from appearing as a distinct ordinary \BLR{}-like \BEL{} component. Absorbed power may instead emerge as diffuse continuum or pseudo-continuum emission \citep{Ferland1988, Rees1989}. This is consistent with dense, optically thick inner-disk expected in the self-consistent $\alpha$-closure framework \citep{naddaf2026}.
The detailed balance between scattering, absorption, and thermal re-emission depends on the density, column density, ionization structure, and geometry. In particular, changes in $U$, $N_{\rm H}$, or covering geometry may produce large changes in the transmitted \EUV/soft-\xray{} continuum, especially when partially ionized layers or ionization fronts are present. Quantifying this behavior calls for dedicated photoionization and transfer modeling. The filter is thus neither a uniform photoabsorbing screen nor a pure electron-scattering mirror, but a stratified structure with opacity varying by depth, frequency, and direction.

This does not require switching between different types of filters. In a stratified disk-atmosphere or wind-base structure, the illuminated surface can be highly ionized and largely scattering-dominated, while deeper or more oblique paths can contain partially ionized, higher-opacity layers. The transmitted continuum is thus dependent on frequency and geometry. Harder \EUV/soft-\xray{} photons, which are most relevant for high-ionization lines (\HIL{s}), are preferentially reduced when the radiation samples partially ionized or higher-column regions, whereas the softer continuum relevant for lower-ionization lines may be less strongly affected. Thus, differential behavior between \HIL{s} and \LIL{s} follows from the stratified opacity structure, not from assigning unrelated properties to the filter in different contexts.

The same structure need not erase reverberation. The classical \BLR{} can still respond to the residual variable ionizing continuum transmitted through lower-opacity surface layers or angular channels. Filtering therefore modifies the reverberation transfer function rather than invalidating reverberation mapping altogether. Depending on the strength, geometry, and variability of the filter, it can reduce line responsivity, introduce continuum--line decorrelation, or produce offsets from the standard radius--luminosity relation. Coherent reverberation is still expected whenever a variable transmitted ionizing component reaches the \BLR{}.

This phenomenological treatment is motivated by observed cases in which the continuum seen by the observer is not necessarily the same as the ionizing continuum driving the \BLR{} response. During the \AGN{\,storm} campaign of NGC\,5548, the \BEL{s} decorrelated from the observed \UV{} continuum, indicating that the \BLR{} responded to a modified ionizing continuum \citep{Kara2021, Homayouni2024}. Wind-based interpretations attribute this behavior to an obscuring or disk-wind component between the central source and the \BLR{}, whose density, ionization state, or covering factor changes the \SED{} incident on the line-emitting gas \citep{Mehdipour2016, Dehghanian2019}. Similar obscuration and anomalous line-response behavior have also been reported in Mrk\,817 \citep{Zaidouni2024}.

This picture is closely related to disk-wind obscurer models developed for NGC\,5548, where the line-of-sight obscurer affects absorption features while the denser wind base or equatorial obscurer modifies the \SED{} incident on the \BLR{}. Such models show that the transmitted continuum can fall into distinct ionization regimes: a highly ionized, nearly transparent state; an intermediate state in which a He ionization front attenuates the harder \EUV/soft-\xray{} photons while allowing substantial Lyman-continuum transmission; and a lower-ionization state in which an H ionization front strongly suppresses the Lyman continuum as well. These regimes provide a natural physical basis for differential filtering of \HIL{s} and \LIL{s}. They also emphasize that $T_{\rm net}$ can depend sensitively on $U$, $N_{\rm H}$, density, and covering geometry, and that the emergent signal from the filter may include transmitted, scattered/reflected, and reprocessed components rather than a single reprocessed continuum \citep[see e.g.,][]{Dehghanian2021}.

The filtering picture is also theoretically motivated. Geometrically thick inner flows, dusty failed outflows, and radiation-driven winds can reduce the ionizing flux reaching larger radii \citep{abramowicz1988,murray1995,proga2000}. In FRADO-like scenarios, the vertical extension of dusty outflows and failed winds provides a natural physical route through which accretion state can regulate the effective ionizing illumination of the \BLR{} \citep{czerny2011, wang_shielding2014, naddaf2021, naddaf2025L}.

The present work therefore requires only that the ionizing continuum can be reduced or reshaped before reaching the line-emitting gas. Predicting the spectrum, luminosity, and angular distribution of the reprocessed emission from the filtering layer is beyond the scope of this Letter. The relevant quantity for line formation is the continuum that survives filtering and reaches the \BLR{} gas. The existence of a finite line-production window then implies that \BEL{} appearance is not controlled by a universal $\lambda_{\rm Edd}$ threshold, but by a coupled condition involving both photon production and transmission.

\subsection{Scope, caveats, and future tests}

The present treatment is intentionally phenomenological and does not provide a full radiative-transfer solution for the filtering layer. We do not calculate the detailed transmitted or reprocessed spectrum, nor test whether the resulting diffuse or pseudo-continuum satisfies source-specific observational constraints. These depend on the density, column density, ionization structure, covering geometry, and velocity field of the filtering medium, and require dedicated photoionization and radiative-transfer calculations.

The reverberation discussion is likewise qualitative. The framework does not predict a detailed transfer function, but emphasizes that filtering need not erase reverberation if a variable transmitted ionizing component still reaches the classical \BLR{}. In that case, filtering may modify responsivity, produce decorrelation episodes, or generate offsets from radius--luminosity relations. The broader implications discussed below should therefore be regarded as physically motivated consequences of the \TR{} framework, to be tested quantitatively in future work.

\subsection{A unified interpretation of \AGN{} phenomenology}

Within this framework, several apparently disconnected aspects of \AGN{} phenomenology can be interpreted as consequences of the same underlying regulation of the ionizing radiation field. The following subsections outline qualitative implications of the \TR{} picture rather than complete source-specific models.

\subsubsection{Weak-line populations}

Low-luminosity \AGN{}s and \LINER{}s naturally fall below the low-accretion boundary of the window. \WLQ{}s occupy the opposite extreme, in which strong filtering suppresses the effective ionizing field reaching the virialized \BLR{} despite high accretion signatures \citep{shemmer2010}. Transitional populations need not be interpreted as distinct classes, but as sources occupying different regions of the same \TR{} parameter space. In this view, weak-line behavior reflects changes in the effective ionizing field rather than a categorical distinction. This naturally unifies low-luminosity \AGN{}s, normal broad-line quasars, and \WLQ{}s within a single radiative-regulation framework.

At low accretion rates, systems such as low-luminosity \AGN{}s and \LINER{}s remain below the ionizing threshold required for \BEL{} production because the intrinsic photon supply is too weak \citep{elitzur2009,cao2010, ho1997, Ho2008, balmaverde2014}. At high accretion rates, by contrast, intrinsic photon production can be large, but strong filtering reduces the fraction of the ionizing continuum that reaches the virialized \BLR{}. In both cases the \BEL-formation condition fails, but for opposite physical regimes. Low- and high-accretion systems therefore occupy opposite ends of the same \TR{} parameter space.

\WLQ{}s, which show high-accretion signatures but unusually faint emission lines \citep{shemmer2010, luo2013, luo2015}, are naturally accounted for within this framework. Recent observations suggest that attenuation in these sources is strongly ionization-dependent, with \HIL{s} suppressed more strongly than lower-ionization lines \citep{cheng2025}. Such behavior is expected if filtering softens the ionizing continuum before it reaches the \BLR{}, reducing the photon supply above the thresholds required for \HIL{} production. \HIL{s} are therefore expected to weaken first, while lower-ionization lines such as \Hbeta{} may remain detectable. The observed \BEL{} visibility is therefore shaped primarily by the wavelength-dependent transmission of the ionizing continuum, which is why filtering need not suppress all \BEL{}s simultaneously.

\subsubsection{Temporal low states and \BLR{} holiday}
In such cases the relevant instantaneous condition is represented by $\mathcal{V}_{t}$ in Eq.~\ref{eq:transmission}, which captures short-timescale variability effects. Thus an \AGN{} may move temporarily into a weak-line state without requiring a secular change in the mass supply through the disk.

This distinction is directly relevant to well-monitored sources such as NGC\,5548, Mrk\,817, NGC\,4151, Mrk\,110, and Mrk\,590, where \BEL{} strengths can vary strongly on observational timescales. In particular, the \BLR{}-holiday behavior observed in NGC\,5548 and Mrk\,817 shows that \BEL{} light curves can decorrelate from the observed \UV{} continuum while strong soft-\xray{} absorption and broad \UV{} absorption indicate intervening material along the ionizing path \citep{kriss2019, Goad2019, Dehghanian2019}. Such behavior is naturally described by time-dependent filtering: the observed \UV{}/optical continuum need not be the same continuum that reaches and drives the \BLR{}. The present model therefore applies both to population-level differences in average \BEL{} strength and to temporary departures, although detailed reverberation behavior requires time-dependent photoionization and transfer modeling.

\subsubsection{\BEL{} profile morphology}
Within this picture, variations in line profile shape can also be understood as a consequence of which regions within the \BLR{} satisfy the line-production condition. If filtering preferentially suppresses the effective ionizing field, then line emission may become dominated, by outflowing components, leading to more pronounced blueshifts and deviations from virial line profiles (see Fig.~\ref{fig:location}). The observed \BEL{} may therefore represent a superposition of a Keplerian core plus a blueshifted wing component \citep{nagoshi2024, coatman2017}.
This requires both a high accretion state capable of launching an outflow and a viewing direction that does not intersect the absorbing outflow cone. Overall, the dominant response may shift between virialized, and outflowing components if present, producing changes in profile shape, and velocity-resolved lag structure. Enhanced blueshifts are therefore a possible outcome, especially for \HIL{s}, and especially at high $\lambda_{\rm Edd}$.

\subsubsection{Breakdown of \BLR{} scaling relations}

High-accretion sources exhibit shorter reverberation lags than expected from standard radius–luminosity relations \citep{kaspi2005, bentz2013}. In this framework, this follows if the \BLR{} responds to the transmitted ionizing field rather than the observed continuum: as filtering increases, the effective ionizing flux decreases, shifting the emitting region inward and reducing the extent of line formation. In extreme cases, such as \WLQ{}s, the transmitted field may fall below the threshold, yielding weak or absent reverberation signals.

More generally, standard \BLR{}-based scaling relations can break down in extreme accretion regimes \citep{du2015, Dupu2019, du_wang2019}, especially if non-virial components contribute significantly to the response-weighted line profile. Virial mass estimators \citep{peterson2004, pancoast2014} and radius–luminosity relations \citep{kaspi2005, bentz2013} assume that the observed continuum traces the ionizing radiation responsible for line formation. If instead the relevant quantity is the transmitted ionizing field, observed luminosity ceases to be a reliable proxy for \BLR{} regulation, particularly when photon trapping and outflows decouple the intrinsic and incident radiation fields. This provides a physically motivated explanation for the failure of these relations in both low-luminosity \AGN{}s and high-accretion quasars. A direct empirical prediction follows: \BEL{} strength and detectability should correlate more strongly with indicators of filtering, such as \xray{} weakness or absorption, than with luminosity alone, especially at high $\lambda_{\rm Edd}$ \citep{ricci2017}.

\subsubsection{Emergence of the Baldwin effect}
This picture naturally accounts for the Baldwin effect, in which emission-line equivalent widths decrease with increasing luminosity \citep{baldwin1977}. Within this framework, increasing accretion rate or luminosity does not necessarily lead to a proportional increase in the ionizing radiation field incident on the BLR, because filtering reduces the transmitted flux at high accretion states.
The equivalent width can be written schematically as
\begin{equation}
\EW \propto \dfrac{F_{\rm BEL}}{F_{\lambda,\mathrm{cont,obs}}}
\propto \dfrac{T_{\rm net}\, \Phi_{\mathrm{int}}}{F_{\lambda,\mathrm{cont,obs}}}
\propto T_{\rm net}\, P,
\end{equation}
where $P\equiv \Phi_{\rm int}/F_{\lambda,\mathrm{cont}}$ encodes the spectral conversion between the observed optical continuum and the ionizing photon flux. In high-accretion regimes, enhanced filtering and spectral softening can both reduce the effective ionizing field relative to the observed continuum, so that both $T_{\rm net}$ and $P$ decline, naturally suppressing the equivalent width. In this sense, the Baldwin effect reflects radiative regulation of the ionizing field rather than a purely local photoionization response \citep{gilbert2003, baskin2004}.

In this framework, the interpretation however is particularly relevant for \HIL{s}, for which the Baldwin effect is strongest. It directly points to radiative regulation of the ionizing field as an important driver of the observed trend \citep{gilbert2003, baskin2004}. The crucial point is that attenuation is strongly ionization dependent. \HIL{s} require photons above relatively high ionization thresholds and are therefore the first to be affected when filtering softens the incident \SED{}. Their luminosities thus increase more slowly than the observed continuum, or may even decline despite increasing intrinsic luminosity, naturally producing a strong negative Baldwin effect. In contrast, \LIL{s} such as \Hbeta{} are powered by softer photons that remain available over a broader range of accretion states; they are therefore comparatively less sensitive to filtering, substantially weakening the Baldwin effect.

\subsubsection{The behavior of $R_{\rm Fe}$}
The framework, by suppressing H{\footnotesize $\beta$} at high $\lambda_{\rm Edd}$, may contribute to the observed $R_{\rm Fe}$ trend along the quasar main sequence \citep{boroson1992, marziani2003b, zamfir2010}. Although a quantitative account of $\mathrm{Fe\,}${\footnotesize II} requires explicit treatment of density, column density, metallicity, and radiative-transfer, optical Fe{\footnotesize\,II} emission is expected to increase with accretion rate and may approach saturation at high values \citep{baldwin2004}. As a result, $R_{\rm Fe}$ increases toward high Eddington ratios, reflecting the combined effect of enhanced Fe{\footnotesize\,II} emission and the progressive suppression of H{\footnotesize $\beta$} due to filtering and spectral softening. 

\subsubsection{\xray{} weakness as a filtering indicator}

An observational consequence follows from the fact that the observed Baldwin effect is governed by the combination of filtering strength and intrinsic \SED{} changes, rather than by luminosity alone. If filtering dominates this combination, as may occur particularly at high $\lambda_{\rm Edd}$ \citep{ricci2017}, then \BEL{} detectability and strength should correlate more strongly with filtering diagnostics, such as \xray{} weakness or absorption, than with continuum luminosity. At fixed $\lambda_{\rm Edd}$, sources with stronger \xray{} weakness, or more negative $\Delta\alpha_{\rm ox}$, are thus expected to exhibit systematically weaker \BEL{}s or a higher probability of \BEL{} absence \citep{leighly2007, ni2018, ni2022}.
$\Delta\alpha_{\rm ox}$ is the offset from the expected $\alpha_{\rm ox}$
\begin{equation}
    \Delta\alpha_{\rm ox} = \alpha_{\rm ox,obs} - \alpha_{\rm ox,exp}(L_{2500})
\end{equation}
where $L_{2500}$ is the \UV{} specific luminosity from empirical relations \citep{Steffen2006, Lusso2016}. In contrast, the intrinsic \xray{-absorption} in low-luminosity, low $\lambda_{\rm Edd}$ sources is very weak \citep{Ho2008}.

\subsubsection{Falsifiable predictions of the framework}
Five concise, testable predictions follow directly from the condition 
$\varphi_{\mathrm{\BEL{}}}\,T_{\mathrm{net}} \ge 1$, and provide observational
discriminants of the framework:\\

\begin{enumerate}

\item \textbf{\BEL{} strength tracks filtering.}

At fixed high $\lambda_{\rm Edd}$, \BEL{} \EW{}s are expected to correlate more tightly with filtering diagnostics than with optical/\UV{} luminosity. This reflects the dependence of line production on the transmitted ionizing field rather than the observed continuum. Observable quantities include \EW(\CIV), \EW(\Hbeta), $\Delta\alpha_{\rm ox}$, and $N_{\rm H}$ \citep[e.g.,][]{ricci2017, chiaraluce2018}.

\item \textbf{Offsets from radius--luminosity relation.}

Highly accreting, \xray{}-weak sources should lie systematically below the canonical \Hbeta{} size relation when plotted against observed $L_{5100}$, reflecting reduced effective ionizing illumination. The scatter should decrease when using a proxy for the transmitted ionizing field (e.g., $\Delta\alpha_{\rm ox}$-corrected luminosity), since the \BLR{} responds to $\Phi_{\mathrm{int}}T_{\mathrm{net}}$ rather than $L_{5100}$ alone \citep[e.g.,][]{korista1997, ferland2020}.

\item \textbf{Line-dependent suppression.}

Ionization dependent nature of filtering implies that \HIL{s} are suppressed before \LIL{s} as the transmitted continuum softens. There should therefore be systematic trends in line ratios such as \EW(\HeII{})/\EW(\Hbeta{}), \EW(\CIV{})/\EW(\MgII{}), and related diagnostics as a function of $\Delta\alpha_{\rm ox}$ or other filtering indicators \citep[e.g.,][]{luo2013, luo2015, cheng2025}.

\item \textbf{Coupled evolution of line profiles.}

As filtering increases at high Eddington ratios, the effective line-emitting region shifts toward non-virial or outflowing components, leading to enhanced profile asymmetry and blueshift (particularly in \HIL{s}), together with reduced contribution of the virialized core. This predicts correlations between C{\footnotesize IV} blueshift, H{\footnotesize $\beta$} core-to-wing structure, and velocity-resolved lag behavior \citep[e.g.,][]{murray1995, coatman2017, nagoshi2024}.

\item \textbf{No universal $\lambda_{\rm Edd}$ threshold.}

There should be no single critical $\lambda_{\rm Edd}$ separating \BEL{} and non-\BEL{} sources. Instead, the line-production boundary depends jointly on $\lambda_{\rm Edd}$, $M_{\bullet}$, and $\eta$ through their impact on both photon production and transmission. This implies that \BEL{} detection fractions should vary across multidimensional parameter space rather than exhibiting a sharp transition with accretion rate alone \citep[e.g.,][]{laor2003, Ho2008, elitzur2009, cao2010}.

\end{enumerate}

\section{Conclusions}\label{sec:conclusion}
We have shown that broad emission-line visibility in \AGN{}s is governed by the ionizing radiation field reaching the \BLR{}, not by the intrinsic luminosity alone. Expressing this field as the product of intrinsic photon production and filtering-induced transmission naturally produces a bounded maximum, and therefore a finite, non-universal line-production window in global parameter space. This provides a unified explanation for the weakening or disappearance of \BEL{}s in both low-accretion systems and high-accretion quasars, and offers a common framework for interpreting a wide range of observed phenomena, including \WLQ{}s, \LINER{}s, Baldwin-effect trends, the $R_{\rm Fe}$ sequence, and the breakdown of standard \BLR{} scaling relations.

A minimal quantitative realization shows that this behavior is generic and does not rely on a specific transmission prescription. The results imply that the key physical variable is the transmitted ionizing field rather than the observed continuum. The framework therefore predicts that \BEL{} strength and detectability should correlate more strongly with filtering diagnostics, such as \xray{} weakness or absorption, than with luminosity alone. The \TR{} picture thus provides a compact, testable basis for understanding broad-line phenomenology across accretion regimes.

\acknowledgments
\section*{acknowledgments}
This work was supported by University of Liege under Special Funds for Research, IPD-STEMA Program.

\newpage

\bibliographystyle{aasjournal}
\bibliography{naddaf}

@ARTICLE{ferland2020,
       author = {{Ferland}, G.~J. and {Done}, C. and {Jin}, C. and {Landt}, H. and {Ward}, M.~J.},
        title = "{State-of-the-art AGN SEDs for photoionization models: BLR predictions confront the observations}",
      journal = {\mnras},
         year = 2020,
        month = jun,
       volume = {494},
       number = {4},
        pages = {5917-5922},
          doi = {10.1093/mnras/staa1207},
archivePrefix = {arXiv},
       eprint = {2004.11873},
 primaryClass = {astro-ph.HE},
       adsurl = {https://ui.adsabs.harvard.edu/abs/2020MNRAS.494.5917F}
}

@ARTICLE{luo2013,
       author = {{Luo}, B. and {Brandt}, W.~N. and {Alexander}, D.~M. and {Harrison}, F.~A. and {Stern}, D. and {Bauer}, F.~E. and {Boggs}, S.~E. and {Christensen}, F.~E. and {Comastri}, A. and {Craig}, W.~W. and {Fabian}, A.~C. and {Farrah}, D. and {Fiore}, F. and {Fuerst}, F. and {Grefenstette}, B.~W. and {Hailey}, C.~J. and {Hickox}, R. and {Madsen}, K.~K. and {Matt}, G. and {Ogle}, P. and {Risaliti}, G. and {Saez}, C. and {Teng}, S.~H. and {Walton}, D.~J. and {Zhang}, W.~W.},
        title = "{Weak Hard X-Ray Emission from Two Broad Absorption Line Quasars Observed with NuSTAR: Compton-thick Absorption or Intrinsic X-Ray Weakness?}",
      journal = {\apj},
         year = 2013,
        month = aug,
       volume = {772},
       number = {2},
          eid = {153},
        pages = {153},
          doi = {10.1088/0004-637X/772/2/153},
archivePrefix = {arXiv},
       eprint = {1306.3500},
 primaryClass = {astro-ph.CO},
       adsurl = {https://ui.adsabs.harvard.edu/abs/2013ApJ...772..153L}
}

@ARTICLE{luo2015,
       author = {{Luo}, B. and {Brandt}, W.~N. and {Hall}, P.~B. and {Wu}, Jianfeng and {Anderson}, S.~F. and {Garmire}, G.~P. and {Gibson}, R.~R. and {Plotkin}, R.~M. and {Richards}, G.~T. and {Schneider}, D.~P. and {Shemmer}, O. and {Shen}, Yue},
        title = "{X-ray Insights into the Nature of PHL 1811 Analogs and Weak Emission-line Quasars: Unification with a Geometrically Thick Accretion Disk?}",
      journal = {\apj},
         year = 2015,
        month = jun,
       volume = {805},
       number = {2},
          eid = {122},
        pages = {122},
          doi = {10.1088/0004-637X/805/2/122},
archivePrefix = {arXiv},
       eprint = {1503.02085},
 primaryClass = {astro-ph.GA},
       adsurl = {https://ui.adsabs.harvard.edu/abs/2015ApJ...805..122L}
}

@ARTICLE{baldwin1977,
       author = {{Baldwin}, Jack A.},
        title = "{Luminosity Indicators in the Spectra of Quasi-Stellar Objects}",
      journal = {\apj},
         year = 1977,
        month = jun,
       volume = {214},
        pages = {679-684},
          doi = {10.1086/155294},
       adsurl = {https://ui.adsabs.harvard.edu/abs/1977ApJ...214..679B}
}

@ARTICLE{cao2010,
       author = {{Cao}, Xinwu},
        title = "{On the Disappearance of the Broad-line Region in Low-luminosity Active Galactic Nuclei: The Role of the Outflows from Advection Dominated Accretion Flows}",
      journal = {\apj},
         year = 2010,
        month = dec,
       volume = {724},
       number = {2},
        pages = {855-860},
          doi = {10.1088/0004-637X/724/2/855},
archivePrefix = {arXiv},
       eprint = {1009.5043},
 primaryClass = {astro-ph.HE},
       adsurl = {https://ui.adsabs.harvard.edu/abs/2010ApJ...724..855C}
}

@ARTICLE{korista1997,
       author = {{Korista}, Kirk and {Ferland}, Gary and {Baldwin}, Jack},
        title = "{Do the Broad Emission Line Clouds See the Same Continuum That We See?}",
      journal = {\apj},
         year = 1997,
        month = oct,
       volume = {487},
       number = {2},
        pages = {555-559},
          doi = {10.1086/304659},
archivePrefix = {arXiv},
       eprint = {astro-ph/9704262},
 primaryClass = {astro-ph},
       adsurl = {https://ui.adsabs.harvard.edu/abs/1997ApJ...487..555K}
}

@ARTICLE{korista1997a,
       author = {{Korista}, Kirk and {Baldwin}, Jack and {Ferland}, Gary and {Verner}, Dima},
        title = "{An Atlas of Computed Equivalent Widths of Quasar Broad Emission Lines}",
      journal = {\apjs},
         year = 1997,
        month = jan,
       volume = {108},
       number = {2},
        pages = {401-415},
          doi = {10.1086/312966},
archivePrefix = {arXiv},
       eprint = {astro-ph/9611220},
 primaryClass = {astro-ph},
       adsurl = {https://ui.adsabs.harvard.edu/abs/1997ApJS..108..401K}
}

@ARTICLE{ni2018,
       author = {{Ni}, Q. and {Brandt}, W.~N. and {Luo}, B. and {Hall}, P.~B. and {Shen}, Yue and {Anderson}, S.~F. and {Plotkin}, R.~M. and {Richards}, Gordon T. and {Schneider}, D.~P. and {Shemmer}, O. and {Wu}, Jianfeng},
        title = "{Connecting the X-ray properties of weak-line and typical quasars: testing for a geometrically thick accretion disk}",
      journal = {\mnras},
         year = 2018,
        month = nov,
       volume = {480},
       number = {4},
        pages = {5184-5202},
          doi = {10.1093/mnras/sty1989},
archivePrefix = {arXiv},
       eprint = {1807.08757},
 primaryClass = {astro-ph.GA},
       adsurl = {https://ui.adsabs.harvard.edu/abs/2018MNRAS.480.5184N}
}

@ARTICLE{ni2022,
       author = {{Ni}, Q. and {Brandt}, W.~N. and {Luo}, B. and {Garmire}, G.~P. and {Hall}, P.~B. and {Plotkin}, R.~M. and {Shemmer}, O. and {Timlin}, J.~D. and {Vito}, F. and {Wu}, J. and {Yi}, W.},
        title = "{Sensitive Chandra coverage of a representative sample of weak-line quasars: revealing the full range of X-ray properties}",
      journal = {\mnras},
         year = 2022,
        month = apr,
       volume = {511},
       number = {4},
        pages = {5251-5264},
          doi = {10.1093/mnras/stac394},
archivePrefix = {arXiv},
       eprint = {2202.05279},
 primaryClass = {astro-ph.GA},
       adsurl = {https://ui.adsabs.harvard.edu/abs/2022MNRAS.511.5251N}
}

@ARTICLE{marziani2003b,
       author = {{Marziani}, Paola and {Zamanov}, Radoslav K. and {Sulentic}, Jack W. and {Calvani}, Massimo},
        title = "{Searching for the physical drivers of eigenvector 1: influence of black hole mass and Eddington ratio}",
      journal = {\mnras},
         year = 2003,
        month = nov,
       volume = {345},
       number = {4},
        pages = {1133-1144},
          doi = {10.1046/j.1365-2966.2003.07033.x},
archivePrefix = {arXiv},
       eprint = {astro-ph/0307367},
 primaryClass = {astro-ph},
       adsurl = {https://ui.adsabs.harvard.edu/abs/2003MNRAS.345.1133M}
}

@ARTICLE{baskin2004,
       author = {{Baskin}, Alexei and {Laor}, Ari},
        title = "{On the origin of the C IV Baldwin effect in active galactic nuclei}",
      journal = {\mnras},
         year = 2004,
        month = may,
       volume = {350},
       number = {2},
        pages = {L31-L35},
          doi = {10.1111/j.1365-2966.2004.07833.x},
archivePrefix = {arXiv},
       eprint = {astro-ph/0403365},
 primaryClass = {astro-ph},
       adsurl = {https://ui.adsabs.harvard.edu/abs/2004MNRAS.350L..31B}
}

@ARTICLE{coatman2017,
       author = {{Coatman}, Liam and {Hewett}, Paul C. and {Banerji}, Manda and {Richards}, Gordon T. and {Hennawi}, Joseph F. and {Prochaska}, J. Xavier},
        title = "{Correcting C IV-based virial black hole masses}",
      journal = {\mnras},
         year = 2017,
        month = feb,
       volume = {465},
       number = {2},
        pages = {2120-2142},
          doi = {10.1093/mnras/stw2797},
archivePrefix = {arXiv},
       eprint = {1610.08977},
 primaryClass = {astro-ph.GA},
       adsurl = {https://ui.adsabs.harvard.edu/abs/2017MNRAS.465.2120C}
}

@ARTICLE{wang_shielding2014,
       author = {{Wang}, Jian-Min and {Qiu}, Jie and {Du}, Pu and {Ho}, Luis C.},
        title = "{Self-shadowing Effects of Slim Accretion Disks in Active Galactic Nuclei: The Diverse Appearance of the Broad-line Region}",
      journal = {\apj},
         year = 2014,
        month = dec,
       volume = {797},
       number = {1},
          eid = {65},
        pages = {65},
          doi = {10.1088/0004-637X/797/1/65},
archivePrefix = {arXiv},
       eprint = {1410.5285},
 primaryClass = {astro-ph.GA},
       adsurl = {https://ui.adsabs.harvard.edu/abs/2014ApJ...797...65W}
}

@ARTICLE{baldwin2004,
       author = {{Baldwin}, J.~A. and {Ferland}, G.~J. and {Korista}, K.~T. and {Hamann}, F. and {LaCluyz{\'e}}, A.},
        title = "{The Origin of Fe II Emission in Active Galactic Nuclei}",
      journal = {\apj},
         year = 2004,
        month = nov,
       volume = {615},
       number = {2},
        pages = {610-624},
          doi = {10.1086/424683},
archivePrefix = {arXiv},
       eprint = {astro-ph/0407404},
 primaryClass = {astro-ph},
       adsurl = {https://ui.adsabs.harvard.edu/abs/2004ApJ...615..610B}
}

@ARTICLE{bentz2013,
       author = {{Bentz}, Misty C. and {Denney}, Kelly D. and {Grier}, Catherine J. and {Barth}, Aaron J. and {Peterson}, Bradley M. and {Vestergaard}, Marianne and {Bennert}, Vardha N. and {Canalizo}, Gabriela and {De Rosa}, Gisella and {Filippenko}, Alexei V. and {Gates}, Elinor L. and {Greene}, Jenny E. and {Li}, Weidong and {Malkan}, Matthew A. and {Pogge}, Richard W. and {Stern}, Daniel and {Treu}, Tommaso and {Woo}, Jong-Hak},
        title = "{The Low-luminosity End of the Radius-Luminosity Relationship for Active Galactic Nuclei}",
      journal = {\apj},
         year = 2013,
        month = apr,
       volume = {767},
       number = {2},
          eid = {149},
        pages = {149},
          doi = {10.1088/0004-637X/767/2/149},
archivePrefix = {arXiv},
       eprint = {1303.1742},
 primaryClass = {astro-ph.CO},
       adsurl = {https://ui.adsabs.harvard.edu/abs/2013ApJ...767..149B}
}

@ARTICLE{murray1995,
       author = {{Murray}, N. and {Chiang}, J. and {Grossman}, S.~A. and {Voit}, G.~M.},
        title = "{Accretion Disk Winds from Active Galactic Nuclei}",
      journal = {\apj},
         year = 1995,
        month = oct,
       volume = {451},
        pages = {498},
          doi = {10.1086/176238},
       adsurl = {https://ui.adsabs.harvard.edu/abs/1995ApJ...451..498M}
}

@ARTICLE{naddaf2025,
       author = {{Naddaf}, M.~H. and {Martinez-Aldama}, M.~L. and {Hutsem{\'e}kers}, D. and {Savic}, D. and {Czerny}, B.},
        title = "{H{\ensuremath{\beta}} line shape and radius-luminosity relation in 2.5D FRADO}",
      journal = {\aap},
         year = 2025,
        month = oct,
       volume = {702},
          eid = {A46},
        pages = {A46},
          doi = {10.1051/0004-6361/202555767},
archivePrefix = {arXiv},
       eprint = {2506.01159},
 primaryClass = {astro-ph.GA},
       adsurl = {https://ui.adsabs.harvard.edu/abs/2025A&A...702A..46N}
}

@ARTICLE{naddaf2025L,
       author = {{Naddaf}, M.~H. and {Mart{\'\i}nez-Aldama}, M.~L. and {Marziani}, P. and {Czerny}, B. and {Hutsem{\'e}kers}, D.},
        title = "{Quasar main sequence unfolded by 2.5D FRADO: Natural expression of Eddington ratio, black hole mass, and inclination}",
      journal = {\aap},
         year = 2025,
        month = oct,
       volume = {702},
          eid = {L13},
        pages = {L13},
          doi = {10.1051/0004-6361/202556852},
archivePrefix = {arXiv},
       eprint = {2510.01092},
 primaryClass = {astro-ph.GA},
       adsurl = {https://ui.adsabs.harvard.edu/abs/2025A&A...702L..13N}
}

@ARTICLE{ho1997,
       author = {{Ho}, Luis C. and {Filippenko}, Alexei V. and {Sargent}, Wallace L.~W.},
        title = "{A Search for ``Dwarf'' Seyfert Nuclei. III. Spectroscopic Parameters and Properties of the Host Galaxies}",
      journal = {\apjs},
         year = 1997,
        month = oct,
       volume = {112},
       number = {2},
        pages = {315-390},
          doi = {10.1086/313041},
archivePrefix = {arXiv},
       eprint = {astro-ph/9704107},
 primaryClass = {astro-ph},
       adsurl = {https://ui.adsabs.harvard.edu/abs/1997ApJS..112..315H}
}

@ARTICLE{balmaverde2014,
       author = {{Balmaverde}, B. and {Capetti}, A.},
        title = "{The HST view of the broad line region in low luminosity AGN}",
      journal = {\aap},
         year = 2014,
        month = mar,
       volume = {563},
          eid = {A119},
        pages = {A119},
          doi = {10.1051/0004-6361/201321989},
archivePrefix = {arXiv},
       eprint = {1401.5242},
 primaryClass = {astro-ph.GA},
       adsurl = {https://ui.adsabs.harvard.edu/abs/2014A&A...563A.119B}
}

@ARTICLE{cheng2025,
       author = {{Cheng}, Xiaoqiang and {Wu}, Jianfeng and {Wu}, Qiaoya},
        title = "{Quest for a Coherent Definition of Weak-line Quasars and Its Physical Implications}",
      journal = {\apj},
         year = 2025,
        month = dec,
       volume = {994},
       number = {2},
          eid = {213},
        pages = {213},
          doi = {10.3847/1538-4357/ae1126},
archivePrefix = {arXiv},
       eprint = {2510.08135},
 primaryClass = {astro-ph.GA},
       adsurl = {https://ui.adsabs.harvard.edu/abs/2025ApJ...994..213C}
}

@ARTICLE{chiaraluce2018,
       author = {{Chiaraluce}, E. and {Vagnetti}, F. and {Tombesi}, F. and {Paolillo}, M.},
        title = "{The X-ray/UV ratio in active galactic nuclei: dispersion and variability}",
      journal = {\aap},
         year = 2018,
        month = nov,
       volume = {619},
          eid = {A95},
        pages = {A95},
          doi = {10.1051/0004-6361/201833631},
archivePrefix = {arXiv},
       eprint = {1808.06964},
 primaryClass = {astro-ph.GA},
       adsurl = {https://ui.adsabs.harvard.edu/abs/2018A&A...619A..95C}
}

@ARTICLE{nagoshi2024,
       author = {{Nagoshi}, Shumpei and {Iwamuro}, Fumihide and {Yamada}, Satoshi and {Ueda}, Yoshihiro and {Oikawa}, Yuto and {Otsuka}, Masaaki and {Isogai}, Keisuke and {Mineshige}, Shin},
        title = "{Probing the origin of the two-component structure of broad-line region by reverberation mapping of an extremely variable quasar}",
      journal = {\mnras},
         year = 2024,
        month = mar,
       volume = {529},
       number = {1},
        pages = {393-408},
          doi = {10.1093/mnras/stae319},
archivePrefix = {arXiv},
       eprint = {2306.13930},
 primaryClass = {astro-ph.GA},
       adsurl = {https://ui.adsabs.harvard.edu/abs/2024MNRAS.529..393N}
}

@ARTICLE{pancoast2014,
       author = {{Pancoast}, Anna and {Brewer}, Brendon J. and {Treu}, Tommaso and {Park}, Daeseong and {Barth}, Aaron J. and {Bentz}, Misty C. and {Woo}, Jong-Hak},
        title = "{Modelling reverberation mapping data - II. Dynamical modelling of the Lick AGN Monitoring Project 2008 data set}",
      journal = {\mnras},
         year = 2014,
        month = dec,
       volume = {445},
       number = {3},
        pages = {3073-3091},
          doi = {10.1093/mnras/stu1419},
archivePrefix = {arXiv},
       eprint = {1311.6475},
 primaryClass = {astro-ph.CO},
       adsurl = {https://ui.adsabs.harvard.edu/abs/2014MNRAS.445.3073P}
}

@ARTICLE{laor2003,
       author = {{Laor}, Ari},
        title = "{On the Nature of Low-Luminosity Narrow-Line Active Galactic Nuclei}",
      journal = {\apj},
         year = 2003,
        month = jun,
       volume = {590},
       number = {1},
        pages = {86-94},
          doi = {10.1086/375008},
archivePrefix = {arXiv},
       eprint = {astro-ph/0302541},
 primaryClass = {astro-ph},
       adsurl = {https://ui.adsabs.harvard.edu/abs/2003ApJ...590...86L}
}

@ARTICLE{czerny2011,
       author = {{Czerny}, B. and {Hryniewicz}, K.},
        title = "{The origin of the broad line region in active galactic nuclei}",
      journal = {\aap},
         year = 2011,
        month = jan,
       volume = {525},
          eid = {L8},
        pages = {L8},
          doi = {10.1051/0004-6361/201016025},
archivePrefix = {arXiv},
       eprint = {1010.6201},
 primaryClass = {astro-ph.CO},
       adsurl = {https://ui.adsabs.harvard.edu/abs/2011A&A...525L...8C}
}

@ARTICLE{ricci2017,
       author = {{Ricci}, Claudio and {Trakhtenbrot}, Benny and {Koss}, Michael J. and {Ueda}, Yoshihiro and {Schawinski}, Kevin and {Oh}, Kyuseok and {Lamperti}, Isabella and {Mushotzky}, Richard and {Treister}, Ezequiel and {Ho}, Luis C. and {Weigel}, Anna and {Bauer}, Franz E. and {Paltani}, Stephane and {Fabian}, Andrew C. and {Xie}, Yanxia and {Gehrels}, Neil},
        title = "{The close environments of accreting massive black holes are shaped by radiative feedback}",
      journal = {\nat},
         year = 2017,
        month = sep,
       volume = {549},
       number = {7673},
        pages = {488-491},
          doi = {10.1038/nature23906},
archivePrefix = {arXiv},
       eprint = {1709.09651},
 primaryClass = {astro-ph.HE},
       adsurl = {https://ui.adsabs.harvard.edu/abs/2017Natur.549..488R}
}

@ARTICLE{shemmer2010,
       author = {{Shemmer}, Ohad and {Trakhtenbrot}, Benny and {Anderson}, Scott F. and {Brandt}, W.~N. and {Diamond-Stanic}, Aleksandar M. and {Fan}, Xiaohui and {Lira}, Paulina and {Netzer}, Hagai and {Plotkin}, Richard M. and {Richards}, Gordon T. and {Schneider}, Donald P. and {Strauss}, Michael A.},
        title = "{Weak Line Quasars at High Redshift: Extremely High Accretion Rates or Anemic Broad-line Regions?}",
      journal = {\apjl},
         year = 2010,
        month = oct,
       volume = {722},
       number = {2},
        pages = {L152-L156},
          doi = {10.1088/2041-8205/722/2/L152},
archivePrefix = {arXiv},
       eprint = {1009.2091},
 primaryClass = {astro-ph.CO},
       adsurl = {https://ui.adsabs.harvard.edu/abs/2010ApJ...722L.152S}
}

@ARTICLE{kaspi2005,
       author = {{Kaspi}, Shai and {Maoz}, Dan and {Netzer}, Hagai and {Peterson}, Bradley M. and {Vestergaard}, Marianne and {Jannuzi}, Buell T.},
        title = "{The Relationship between Luminosity and Broad-Line Region Size in Active Galactic Nuclei}",
      journal = {\apj},
         year = 2005,
        month = aug,
       volume = {629},
       number = {1},
        pages = {61-71},
          doi = {10.1086/431275},
archivePrefix = {arXiv},
       eprint = {astro-ph/0504484},
 primaryClass = {astro-ph},
       adsurl = {https://ui.adsabs.harvard.edu/abs/2005ApJ...629...61K}
}

@ARTICLE{leighly2007,
       author = {{Leighly}, Karen M. and {Halpern}, Jules P. and {Jenkins}, Edward B. and {Casebeer}, Darrin},
        title = "{The Intrinsically X-Ray-weak Quasar PHL 1811. II. Optical and UV Spectra and Analysis}",
      journal = {\apjs},
         year = 2007,
        month = nov,
       volume = {173},
       number = {1},
        pages = {1-36},
          doi = {10.1086/519768},
archivePrefix = {arXiv},
       eprint = {0705.0940},
 primaryClass = {astro-ph},
       adsurl = {https://ui.adsabs.harvard.edu/abs/2007ApJS..173....1L}
}

@ARTICLE{zamfir2010,
       author = {{Zamfir}, S. and {Sulentic}, J.~W. and {Marziani}, P. and {Dultzin}, D.},
        title = "{Detailed characterization of H{\ensuremath{\beta}} emission line profile in low-z SDSS quasars}",
      journal = {\mnras},
         year = 2010,
        month = apr,
       volume = {403},
       number = {4},
        pages = {1759-1786},
          doi = {10.1111/j.1365-2966.2009.16236.x},
archivePrefix = {arXiv},
       eprint = {0912.4306},
 primaryClass = {astro-ph.CO},
       adsurl = {https://ui.adsabs.harvard.edu/abs/2010MNRAS.403.1759Z}
}

@ARTICLE{peterson2004,
       author = {{Peterson}, B.~M. and {Ferrarese}, L. and {Gilbert}, K.~M. and
         {Kaspi}, S. and {Malkan}, M.~A. and {Maoz}, D. and {Merritt}, D. and
         {Netzer}, H. and {Onken}, C.~A. and {Pogge}, R.~W. and
         {Vestergaard}, M. and {Wandel}, A.},
        title = "{Central Masses and Broad-Line Region Sizes of Active Galactic Nuclei. II. A Homogeneous Analysis of a Large Reverberation-Mapping Database}",
      journal = {ApJ},
         year = "2004",
        month = "Oct",
       volume = {613},
       number = {2},
        pages = {682-699},
          doi = {10.1086/423269},
archivePrefix = {arXiv},
       eprint = {astro-ph/0407299},
 primaryClass = {astro-ph},
       adsurl = {https://ui.adsabs.harvard.edu/abs/2004ApJ...613..682P}
}

@ARTICLE{naddaf2021,
       author = {{Naddaf}, Mohammad-Hassan and {Czerny}, Bo{\.z}ena and {Szczerba}, Ryszard},
        title = "{The Picture of BLR in 2.5D FRADO: Dynamics and Geometry}",
      journal = {\apj},
         year = 2021,
        month = oct,
       volume = {920},
       number = {1},
          eid = {30},
        pages = {30},
          doi = {10.3847/1538-4357/ac139d},
archivePrefix = {arXiv},
       eprint = {2102.00336},
 primaryClass = {astro-ph.GA},
       adsurl = {https://ui.adsabs.harvard.edu/abs/2021ApJ...920...30N}
}

@ARTICLE{boroson1992,
       author = {{Boroson}, Todd A. and {Green}, Richard F.},
        title = "{The Emission-Line Properties of Low-Redshift Quasi-stellar Objects}",
      journal = {\apjs},
         year = 1992,
        month = may,
       volume = {80},
        pages = {109},
          doi = {10.1086/191661},
       adsurl = {https://ui.adsabs.harvard.edu/abs/1992ApJS...80..109B}
}

@ARTICLE{antonucci1993,
   author = {{Antonucci}, R.},
    title = "{Unified models for active galactic nuclei and quasars}",
  journal = {ARA\&A},
     year = 1993,
   volume = 31,
    pages = {473-521},
      doi = {10.1146/annurev.aa.31.090193.002353},
   adsurl = {https://ui.adsabs.harvard.edu/abs/1993ARA%26A..31..473A}
}

@ARTICLE{Dupu2019,
       author = {{Du}, Pu and {Wang}, Jian-Min},
        title = "{The Radius-Luminosity Relationship Depends on Optical Spectra in Active Galactic Nuclei}",
      journal = {\apj},
         year = 2019,
        month = nov,
       volume = {886},
       number = {1},
          eid = {42},
        pages = {42},
          doi = {10.3847/1538-4357/ab4908},
archivePrefix = {arXiv},
       eprint = {1909.06735},
 primaryClass = {astro-ph.GA},
       adsurl = {https://ui.adsabs.harvard.edu/abs/2019ApJ...886...42D}
}

@ARTICLE{proga2000,
       author = {{Proga}, Daniel and {Stone}, James M. and {Kallman}, Timothy R.},
        title = "{Dynamics of Line-driven Disk Winds in Active Galactic Nuclei}",
      journal = {\apj},
         year = 2000,
        month = nov,
       volume = {543},
       number = {2},
        pages = {686-696},
          doi = {10.1086/317154},
archivePrefix = {arXiv},
       eprint = {astro-ph/0005315},
 primaryClass = {astro-ph},
       adsurl = {https://ui.adsabs.harvard.edu/abs/2000ApJ...543..686P}
}

@ARTICLE{elitzur2009,
       author = {{Elitzur}, Moshe and {Ho}, Luis C.},
        title = "{On the Disappearance of the Broad-Line Region in Low-Luminosity Active Galactic Nuclei}",
      journal = {\apjl},
         year = 2009,
        month = aug,
       volume = {701},
       number = {2},
        pages = {L91-L94},
          doi = {10.1088/0004-637X/701/2/L91},
archivePrefix = {arXiv},
       eprint = {0907.3752},
 primaryClass = {astro-ph.CO},
       adsurl = {https://ui.adsabs.harvard.edu/abs/2009ApJ...701L..91E}
}

@ARTICLE{du2015,
       author = {{Du}, Pu and {Hu}, Chen and {Lu}, Kai-Xing and {Huang}, Ying-Ke and {Cheng}, Cheng and {Qiu}, Jie and {Li}, Yan-Rong and {Zhang}, Yang-Wei and {Fan}, Xu-Liang and {Bai}, Jin-Ming and {Bian}, Wei-Hao and {Yuan}, Ye-Fei and {Kaspi}, Shai and {Ho}, Luis C. and {Netzer}, Hagai and {Wang}, Jian-Min and {SEAMBH Collaboration}},
        title = "{Supermassive Black Holes with High Accretion Rates in Active Galactic Nuclei. IV. H{\ensuremath{\beta}} Time Lags and Implications for Super-Eddington Accretion}",
      journal = {\apj},
         year = 2015,
        month = jun,
       volume = {806},
       number = {1},
          eid = {22},
        pages = {22},
          doi = {10.1088/0004-637X/806/1/22},
archivePrefix = {arXiv},
       eprint = {1504.01844},
 primaryClass = {astro-ph.GA},
       adsurl = {https://ui.adsabs.harvard.edu/abs/2015ApJ...806...22D}
}

@ARTICLE{kriss2019,
       author = {{Kriss}, G.~A. and {De Rosa}, G. and {Ely}, J. and {Peterson}, B.~M. and {Kaastra}, J. and {Mehdipour}, M. and {Ferland}, G.~J. and {Dehghanian}, M. and {Mathur}, S. and {Edelson}, R. and {Korista}, K.~T. and {Arav}, N. and {Barth}, A.~J. and {Bentz}, M.~C. and {Brandt}, W.~N. and {Crenshaw}, D.~M. and {Dalla Bont{\`a}}, E. and {Denney}, K.~D. and {Done}, C. and {Eracleous}, M. and {Fausnaugh}, M.~M. and {Gardner}, E. and {Goad}, M.~R. and {Grier}, C.~J. and {Horne}, Keith and {Kochanek}, C.~S. and {McHardy}, I.~M. and {Netzer}, H. and {Pancoast}, A. and {Pei}, L. and {Pogge}, R.~W. and {Proga}, D. and {Silva}, C. and {Tejos}, N. and {Vestergaard}, M. and {Adams}, S.~M. and {Anderson}, M.~D. and {Ar{\'e}valo}, P. and {Beatty}, T.~G. and {Behar}, E. and {Bennert}, V.~N. and {Bianchi}, S. and {Bigley}, A. and {Bisogni}, S. and {Boissay-Malaquin}, R. and {Borman}, G.~A. and {Bottorff}, M.~C. and {Breeveld}, A.~A. and {Brotherton}, M. and {Brown}, J.~E. and {Brown}, J.~S. and {Cackett}, E.~M. and {Canalizo}, G. and {Cappi}, M. and {Carini}, M.~T. and {Clubb}, K.~I. and {Comerford}, J.~M. and {Coker}, C.~T. and {Corsini}, E.~M. and {Costantini}, E. and {Croft}, S. and {Croxall}, K.~V. and {Deason}, A.~J. and {De Lorenzo-C{\'a}ceres}, A. and {De Marco}, B. and {Dietrich}, M. and {Di Gesu}, L. and {Ebrero}, J. and {Evans}, P.~A. and {Filippenko}, A.~V. and {Flatland}, K. and {Gates}, E.~L. and {Gehrels}, N. and {Geier}, S. and {Gelbord}, J.~M. and {Gonzalez}, L. and {Gorjian}, V. and {Grupe}, D. and {Gupta}, A. and {Hall}, P.~B. and {Henderson}, C.~B. and {Hicks}, S. and {Holmbeck}, E. and {Holoien}, T.~W. -S. and {Hutchison}, T.~A. and {Im}, M. and {Jensen}, J.~J. and {Johnson}, C.~A. and {Joner}, M.~D. and {Kaspi}, S. and {Kelly}, B.~C. and {Kelly}, P.~L. and {Kennea}, J.~A. and {Kim}, M. and {Kim}, S.~C. and {Kim}, S.~Y. and {King}, A. and {Klimanov}, S.~A. and {Krongold}, Y. and {Lau}, M.~W. and {Lee}, J.~C. and {Leonard}, D.~C. and {Li}, Miao and {Lira}, P. and {Lochhaas}, C. and {Ma}, Zhiyuan and {MacInnis}, F. and {Malkan}, M.~A. and {Manne-Nicholas}, E.~R. and {Matt}, G. and {Mauerhan}, J.~C. and {McGurk}, R. and {Montuori}, C. and {Morelli}, L. and {Mosquera}, A. and {Mudd}, D. and {M{\"u}ller-S{\'a}nchez}, F. and {Nazarov}, S.~V. and {Norris}, R.~P. and {Nousek}, J.~A. and {Nguyen}, M.~L. and {Ochner}, P. and {Okhmat}, D.~N. and {Paltani}, S. and {Parks}, J.~R. and {Pinto}, C. and {Pizzella}, A. and {Poleski}, R. and {Ponti}, G. and {Pott}, J. -U. and {Rafter}, S.~E. and {Rix}, H. -W. and {Runnoe}, J. and {Saylor}, D.~A. and {Schimoia}, J.~S. and {Schn{\"u}lle}, K. and {Scott}, B. and {Sergeev}, S.~G. and {Shappee}, B.~J. and {Shivvers}, I. and {Siegel}, M. and {Simonian}, G.~V. and {Siviero}, A. and {Skielboe}, A. and {Somers}, G. and {Spencer}, M. and {Starkey}, D. and {Stevens}, D.~J. and {Sung}, H. -I. and {Tayar}, J. and {Teems}, K.~G. and {Treu}, T. and {Turner}, C.~S. and {Uttley}, P. and {. Van Saders}, J. and {Vican}, L. and {Villforth}, C. and {Villanueva}, Jr., S. and {Walton}, D.~J. and {Waters}, T. and {Weiss}, Y. and {Woo}, J. -H. and {Yan}, H. and {Yuk}, H. and {Zheng}, W. and {Zhu}, W. and {Zu}, Y.},
        title = "{Space Telescope and Optical Reverberation Mapping Project. VIII. Time Variability of Emission and Absorption in NGC 5548 Based on Modeling the Ultraviolet Spectrum}",
      journal = {\apj},
         year = 2019,
        month = aug,
       volume = {881},
       number = {2},
          eid = {153},
        pages = {153},
          doi = {10.3847/1538-4357/ab3049},
archivePrefix = {arXiv},
       eprint = {1907.03874},
 primaryClass = {astro-ph.GA},
       adsurl = {https://ui.adsabs.harvard.edu/abs/2019ApJ...881..153K}
}

@ARTICLE{gilbert2003,
       author = {{Gilbert}, Karoline M. and {Peterson}, Bradley M.},
        title = "{An Intrinsic Baldwin Effect in the H{\ensuremath{\beta}} Broad Emission Line in the Spectrum of NGC 5548}",
      journal = {\apj},
         year = 2003,
        month = apr,
       volume = {587},
       number = {1},
        pages = {123-127},
          doi = {10.1086/368112},
archivePrefix = {arXiv},
       eprint = {astro-ph/0212379},
 primaryClass = {astro-ph},
       adsurl = {https://ui.adsabs.harvard.edu/abs/2003ApJ...587..123G}
}

@ARTICLE{korista2004,
       author = {{Korista}, Kirk T. and {Goad}, Michael R.},
        title = "{What the Optical Recombination Lines Can Tell Us about the Broad-Line Regions of Active Galactic Nuclei}",
      journal = {\apj},
         year = 2004,
        month = may,
       volume = {606},
       number = {2},
        pages = {749-762},
          doi = {10.1086/383193},
archivePrefix = {arXiv},
       eprint = {astro-ph/0402506},
 primaryClass = {astro-ph},
       adsurl = {https://ui.adsabs.harvard.edu/abs/2004ApJ...606..749K}
}

@ARTICLE{du_wang2019,
       author = {{Du}, Pu and {Wang}, Jian-Min},
        title = "{The Radius-Luminosity Relationship Depends on Optical Spectra in Active Galactic Nuclei}",
      journal = {\apj},
         year = 2019,
        month = nov,
       volume = {886},
       number = {1},
          eid = {42},
        pages = {42},
          doi = {10.3847/1538-4357/ab4908},
archivePrefix = {arXiv},
       eprint = {1909.06735},
 primaryClass = {astro-ph.GA},
       adsurl = {https://ui.adsabs.harvard.edu/abs/2019ApJ...886...42D}
}

@ARTICLE{abramowicz1988,
       author = {{Abramowicz}, M.~A. and {Czerny}, B. and {Lasota}, J.~P. and {Szuszkiewicz}, E.},
        title = "{Slim Accretion Disks}",
      journal = {\apj},
         year = 1988,
        month = sep,
       volume = {332},
        pages = {646},
          doi = {10.1086/166683},
       adsurl = {https://ui.adsabs.harvard.edu/abs/1988ApJ...332..646A}
}

@ARTICLE{Ho2008,
       author = {{Ho}, L.~C.},
        title = "{Nuclear activity in nearby galaxies.}",
      journal = {\araa},
         year = 2008,
        month = sep,
       volume = {46},
        pages = {475-539},
          doi = {10.1146/annurev.astro.45.051806.110546},
archivePrefix = {arXiv},
       eprint = {0803.2268},
 primaryClass = {astro-ph},
       adsurl = {https://ui.adsabs.harvard.edu/abs/2008ARA&A..46..475H}
}

@ARTICLE{Dehghanian2019,
       author = {{Dehghanian}, M. and {Ferland}, G.~J. and {Peterson}, B.~M. and {Kriss}, G.~A. and {Korista}, K.~T. and {Chatzikos}, M. and {Guzm{\'a}n}, F. and {Arav}, N. and {De Rosa}, G. and {Goad}, M.~R. and {Mehdipour}, M. and {van Hoof}, P.~A.~M.},
        title = "{A Wind-based Unification Model for NGC 5548: Spectral Holidays, Nondisk Emission, and Implications for Changing-look Quasars}",
      journal = {\apjl},
         year = 2019,
        month = sep,
       volume = {882},
       number = {2},
          eid = {L30},
        pages = {L30},
          doi = {10.3847/2041-8213/ab3d41},
archivePrefix = {arXiv},
       eprint = {1908.07686},
 primaryClass = {astro-ph.GA},
       adsurl = {https://ui.adsabs.harvard.edu/abs/2019ApJ...882L..30D}
}

@ARTICLE{Goad2019,
       author = {{Goad}, M.~R. and {Knigge}, C. and {Korista}, K.~T. and {Cackett}, E. and {Horne}, K. and {Starkey}, D.~A. and {Peterson}, B.~M. and {De Rosa}, G. and {Kriss}, G.~A. and {Edelson}, R. and {Fausnaugh}, M.},
        title = "{Anomalous behaviour of the UV-optical continuum bands in NGC 5548}",
      journal = {\mnras},
         year = 2019,
        month = jul,
       volume = {486},
       number = {4},
        pages = {5362-5376},
          doi = {10.1093/mnras/stz1186},
archivePrefix = {arXiv},
       eprint = {1904.12588},
 primaryClass = {astro-ph.GA},
       adsurl = {https://ui.adsabs.harvard.edu/abs/2019MNRAS.486.5362G}
}

@ARTICLE{Homayouni2024,
       author = {{Homayouni}, Y. and {Kriss}, Gerard A. and {De Rosa}, Gisella and {Plesha}, Rachel and {Cackett}, Edward M. and {Goad}, Michael R. and {Korista}, Kirk T. and {Horne}, Keith and {Fischer}, Travis and {Waters}, Tim and {Barth}, Aaron J. and {Kara}, Erin A. and {Landt}, Hermine and {Arav}, Nahum and {Boizelle}, Benjamin D. and {Bentz}, Misty C. and {Brotherton}, Michael S. and {Chelouche}, Doron and {Dalla Bont{\`a}}, Elena and {Dehghanian}, Maryam and {Du}, Pu and {Ferland}, Gary J. and {Fian}, Carina and {Gelbord}, Jonathan and {Grier}, Catherine J. and {Hall}, Patrick B. and {Hu}, Chen and {Ili{\'c}}, Dragana and {Joner}, Michael D. and {Kaastra}, Jelle and {Kaspi}, Shai and {Kova{\v{c}}evi{\'c}}, Andjelka B. and {Kynoch}, Daniel and {Li}, Yan-Rong and {Mehdipour}, Missagh and {Miller}, Jake A. and {Mitchell}, Jake and {Montano}, John and {Netzer}, Hagai and {Neustadt}, J.~M.~M. and {Partington}, Ethan and {Popovi{\'c}}, Luka {\v{C}}. and {Proga}, Daniel and {Storchi-Bergmann}, Thaisa and {Sanmartim}, David and {Siebert}, Matthew R. and {Treu}, Tommaso and {Vestergaard}, Marianne and {Wang}, Jian-Min and {Ward}, Martin J. and {Zaidouni}, Fatima and {Zu}, Ying},
        title = "{AGN STORM 2. V. Anomalous Behavior of the C IV Light Curve of Mrk 817}",
      journal = {\apj},
         year = 2024,
        month = mar,
       volume = {963},
       number = {2},
          eid = {123},
        pages = {123},
          doi = {10.3847/1538-4357/ad1be4},
archivePrefix = {arXiv},
       eprint = {2308.00742},
 primaryClass = {astro-ph.GA},
       adsurl = {https://ui.adsabs.harvard.edu/abs/2024ApJ...963..123H}
}

@ARTICLE{Kara2021,
       author = {{Kara}, Erin and {Mehdipour}, Missagh and {Kriss}, Gerard A. and {Cackett}, Edward M. and {Arav}, Nahum and {Barth}, Aaron J. and {Byun}, Doyee and {Brotherton}, Michael S. and {De Rosa}, Gisella and {Gelbord}, Jonathan and {Hern{\'a}ndez Santisteban}, Juan V. and {Hu}, Chen and {Kaastra}, Jelle and {Landt}, Hermine and {Li}, Yan-Rong and {Miller}, Jake A. and {Montano}, John and {Partington}, Ethan and {Aceituno}, Jes{\'u}s and {Bai}, Jin-Ming and {Bao}, Dongwei and {Bentz}, Misty C. and {Brink}, Thomas G. and {Chelouche}, Doron and {Chen}, Yong-Jie and {Colmenero}, Encarni Romero and {Bont{\`a}}, Elena Dalla and {Dehghanian}, Maryam and {Du}, Pu and {Edelson}, Rick and {Ferland}, Gary J. and {Ferrarese}, Laura and {Fian}, Carina and {Filippenko}, Alexei V. and {Fischer}, Travis and {Goad}, Michael R. and {Gonz{\'a}lez Buitrago}, Diego H. and {Gorjian}, Varoujan and {Grier}, Catherine J. and {Guo}, Wei-Jian and {Hall}, Patrick B. and {Ho}, Luis C. and {Homayouni}, Y. and {Horne}, Keith and {Ili{\'c}}, Dragana and {Jiang}, Bo-Wei and {Joner}, Michael D. and {Kaspi}, Shai and {Kochanek}, Christopher S. and {Korista}, Kirk T. and {Kynoch}, Daniel and {Li}, Sha-Sha and {Liu}, Jun-Rong and {McHardy}, Ian M. and {McLane}, Jacob N. and {Mitchell}, Jake A.~J. and {Netzer}, Hagai and {Olson}, Kianna A. and {Pogge}, Richard W. and {Popovi{\'c}}, Luka C̆. and {Proga}, Daniel and {Storchi-Bergmann}, Thaisa and {Strasburger}, Erika and {Treu}, Tommaso and {Vestergaard}, Marianne and {Wang}, Jian-Min and {Ward}, Martin J. and {Waters}, Tim and {Williams}, Peter R. and {Yang}, Sen and {Yao}, Zhu-Heng and {Zastrocky}, Theodora E. and {Zhai}, Shuo and {Zu}, Ying},
        title = "{AGN STORM 2. I. First results: A Change in the Weather of Mrk 817}",
      journal = {\apj},
         year = 2021,
        month = dec,
       volume = {922},
       number = {2},
          eid = {151},
        pages = {151},
          doi = {10.3847/1538-4357/ac2159},
archivePrefix = {arXiv},
       eprint = {2105.05840},
 primaryClass = {astro-ph.HE},
       adsurl = {https://ui.adsabs.harvard.edu/abs/2021ApJ...922..151K}
}

@ARTICLE{Ferland1988,
       author = {{Ferland}, G.~J. and {Rees}, M.~J.},
        title = "{Radiative Equilibrium of High-Density Clouds, with Application to Active Galactic Nucleus Continua}",
      journal = {\apj},
         year = 1988,
        month = sep,
       volume = {332},
        pages = {141},
          doi = {10.1086/166639},
       adsurl = {https://ui.adsabs.harvard.edu/abs/1988ApJ...332..141F}
}

@ARTICLE{Zaidouni2024,
       author = {{Zaidouni}, Fatima and {Kara}, Erin and {Kosec}, Peter and {Mehdipour}, Missagh and {Rogantini}, Daniele and {Kriss}, Gerard A. and {Behar}, Ehud and {Kaastra}, Jelle and {Barth}, Aaron J. and {Cackett}, Edward M. and {De Rosa}, Gisella and {Homayouni}, Yasaman and {Horne}, Keith and {Landt}, Hermine and {Arav}, Nahum and {Bentz}, Misty C. and {Brotherton}, Michael S. and {Dalla Bont{\`a}}, Elena and {Dehghanian}, Maryam and {Ferland}, Gary J. and {Fian}, Carina and {Gelbord}, Jonathan and {Goad}, Michael R. and {Gonz{\'a}lez Buitrago}, Diego H. and {Grier}, Catherine J. and {Hall}, Patrick B. and {Hu}, Chen and {Ili{\'c}}, Dragana and {Kaspi}, Shai and {Kochanek}, Christopher S. and {Kova{\v{c}}evi{\'c}}, Andjelka B. and {Kynoch}, Daniel and {Lewin}, Collin and {Montano}, John and {Netzer}, Hagai and {Neustadt}, Jack M.~M. and {Panagiotou}, Christos and {Partington}, Ethan R. and {Plesha}, Rachel and {Popovi{\'c}}, Luka {\v{C}}. and {Proga}, Daniel and {Storchi-Bergmann}, Thaisa and {Sanmartim}, David and {Siebert}, Matthew R. and {Signorini}, Matilde and {Vestergaard}, Marianne and {Waters}, Tim and {Zu}, Ying},
        title = "{AGN STORM 2. IX. Studying the Dynamics of the Ionized Obscurer in Mrk 817 with High-resolution X-Ray Spectroscopy}",
      journal = {\apj},
         year = 2024,
        month = oct,
       volume = {974},
       number = {1},
          eid = {91},
        pages = {91},
          doi = {10.3847/1538-4357/ad6771},
archivePrefix = {arXiv},
       eprint = {2406.17061},
 primaryClass = {astro-ph.HE},
       adsurl = {https://ui.adsabs.harvard.edu/abs/2024ApJ...974...91Z}
}

@ARTICLE{Mehdipour2016,
       author = {{Mehdipour}, M. and {Kaastra}, J.~S. and {Kriss}, G.~A. and {Cappi}, M. and {Petrucci}, P.-O. and {De Marco}, B. and {Ponti}, G. and {Steenbrugge}, K.~C. and {Behar}, E. and {Bianchi}, S. and {Branduardi-Raymont}, G. and {Costantini}, E. and {Ebrero}, J. and {Di Gesu}, L. and {Matt}, G. and {Paltani}, S. and {Peterson}, B.~M. and {Ursini}, F. and {Whewell}, M.},
        title = "{Anatomy of the AGN in NGC 5548. VII. Swift study of obscuration and broadband continuum variability}",
      journal = {\aap},
         year = 2016,
        month = apr,
       volume = {588},
          eid = {A139},
        pages = {A139},
          doi = {10.1051/0004-6361/201527729},
archivePrefix = {arXiv},
       eprint = {1602.03017},
 primaryClass = {astro-ph.HE},
       adsurl = {https://ui.adsabs.harvard.edu/abs/2016A&A...588A.139M}
}

@ARTICLE{Rees1989,
       author = {{Rees}, M.~J. and {Netzer}, Hagai and {Ferland}, G.~J.},
        title = "{Small Dense Broad-Line Regions in Active Nuclei}",
      journal = {\apj},
         year = 1989,
        month = dec,
       volume = {347},
        pages = {640},
          doi = {10.1086/168155},
       adsurl = {https://ui.adsabs.harvard.edu/abs/1989ApJ...347..640R}
}

@ARTICLE{Steffen2006,
       author = {{Steffen}, A.~T. and {Strateva}, I. and {Brandt}, W.~N. and {Alexander}, D.~M. and {Koekemoer}, A.~M. and {Lehmer}, B.~D. and {Schneider}, D.~P. and {Vignali}, C.},
        title = "{The X-Ray-to-Optical Properties of Optically Selected Active Galaxies over Wide Luminosity and Redshift Ranges}",
      journal = {\aj},
         year = 2006,
        month = jun,
       volume = {131},
       number = {6},
        pages = {2826-2842},
          doi = {10.1086/503627},
archivePrefix = {arXiv},
       eprint = {astro-ph/0602407},
 primaryClass = {astro-ph},
       adsurl = {https://ui.adsabs.harvard.edu/abs/2006AJ....131.2826S}
}

@ARTICLE{Lusso2016,
       author = {{Lusso}, E. and {Risaliti}, G.},
        title = "{The Tight Relation between X-Ray and Ultraviolet Luminosity of Quasars}",
      journal = {\apj},
         year = 2016,
        month = mar,
       volume = {819},
       number = {2},
          eid = {154},
        pages = {154},
          doi = {10.3847/0004-637X/819/2/154},
archivePrefix = {arXiv},
       eprint = {1602.01090},
 primaryClass = {astro-ph.GA},
       adsurl = {https://ui.adsabs.harvard.edu/abs/2016ApJ...819..154L}
}

@misc{naddaf2026,
      title={Radiation-pressure instability is an artifact of constant-$\alpha$ closure}, 
      author={M. H. Naddaf and M. Ghasemnezhad and H. Ghanbarnejad and D. Hutsemékers and B. Czerny},
      year={2026},
      eprint={2606.31998},
      archivePrefix={arXiv},
      primaryClass={astro-ph.HE},
      url={https://arxiv.org/abs/2606.31998}, 
}

@ARTICLE{Dehghanian2021,
       author = {{Dehghanian}, M. and {Ferland}, G.~J. and {Peterson}, B.~M. and {Kriss}, G.~A. and {Korista}, K.~T. and {Goad}, M.~R. and {Chatzikos}, M. and {Bentz}, M.~C. and {Guzm{\'a}n}, F. and {Mehdipour}, M. and {De Rosa}, G.},
        title = "{Space Telescope and Optical Reverberation Mapping Project. XIII. An Atlas of UV and X-Ray Spectroscopic Signatures of the Disk Wind in NGC 5548}",
      journal = {\apj},
         year = 2021,
        month = jan,
       volume = {906},
       number = {1},
          eid = {14},
        pages = {14},
          doi = {10.3847/1538-4357/abcb91},
archivePrefix = {arXiv},
       eprint = {2011.09056},
 primaryClass = {astro-ph.GA},
       adsurl = {https://ui.adsabs.harvard.edu/abs/2021ApJ...906...14D}
}

\end{document}